\documentclass[11pt,tightenlines,british,aps,pra,nofootinbib]{revtex4-1}
\usepackage[top=1in, bottom=1in, left=1in, right=1in]{geometry}
\usepackage{amsmath}
\usepackage{amsfonts}
\usepackage{graphicx}
\usepackage{amssymb}
\usepackage{textcomp}
\usepackage{suffix}
\usepackage{mathtools}
\usepackage{mathrsfs}
\usepackage{titlesec}
\usepackage{hyperref}
\usepackage{bm}
\usepackage{dcolumn}
\usepackage[nottoc,numbib]{tocbibind}
\usepackage[titletoc]{appendix}
\graphicspath{ {"home/bogdan/Documents/Books/Physics/Part III/Essay/LaTeX/arXiv"} }
\usepackage{adjustbox}
\usepackage{blkarray}
\usepackage{subfig}
\usepackage{tabularx}
\usepackage[labelfont={small,bf}]{caption}
\usepackage{multirow}
\usepackage{cleveref}
\usepackage{enumitem}
\usepackage{color}
\usepackage{tensor}

\usepackage{fancyhdr}
\usepackage{xargs,lipsum,caption,changepage,ifthen}

\newcommandx{\mycaptionminipage}[3][3=c,usedefault]{%
    \begin{minipage}[#3]{#1}%
        \ifthenelse{\equal{#3}{b}}{\captionsetup{aboveskip=0pt}}{}
        \ifthenelse{\equal{#3}{t}}{\captionsetup{belowskip=0pt}}{}
        \vspace{0pt}\centering\captionsetup{width=\textwidth} 
        #2%
    \end{minipage}%
}%
\newcommandx{\mysidecaption}[4][4=c,usedefault]{%
    \checkoddpage%
    \ifoddpage%
        \mycaptionminipage{\dimexpr\linewidth-#1\linewidth-\intextsep\relax}{#3}[#4]%
        \hfill%
        \mycaptionminipage{#1\linewidth}{#2}[#4]%
    \else%
        \mycaptionminipage{#1\linewidth}{#2}[#4]%
        \hfill%
        \mycaptionminipage{\dimexpr\linewidth-#1\linewidth-\intextsep\relax}{#3}[#4]%
    \fi%
}%

\begin{document}
\setlength{\tabcolsep}{4pt}
\setlength{\intextsep}{1.0\baselineskip}
\title{\Huge\bf{Superradiant instability in AdS}}
\author{\LARGE{Bogdan Ganchev\\ Set By: Jorge Santos}\\
Part III of the Mathematical Tripos, Cambridge\\2015-2016\\}
\vspace*{15mm}
\begin{abstract}
\vspace*{7mm}
\begin{center}
\textbf{Abstract}
\end{center}
The phenomenon of superradiance in the context of asymptotically global AdS spacetimes is investigated with particular accent on its effect on the stability of the systems under consideration. To this end, the concept of an asymptotically AdS spacetime is explained, together with its implications on the boundary conditions at $\mathcal{I}$, as well as the Newman-Penrose-Teukolsky formalism, whereby the Teukolsky master equation in a most general form for Kerr-AdS is given. Furthermore, work done in the cases of RN-AdS and Kerr-AdS is laid out in a concise manner, putting emphasis on the important steps taken in determining the endpoint of the superradiant instability in the two configurations. For the former this turns out to be a black hole with reduced charge and a static charged scalar condensate around it, whereas for the latter two of the more probable outcomes are presented, both of which imply a violation of one of the cosmic censorships.
\end{abstract}
\maketitle
\clearpage
\newpage
\tableofcontents
\section{Structural overview}
This work focusses on the recent developments in the area of superradiant scattering, primarily in asymptotically global Anti-de Sitter spacetimes, aiming at reviewing what the authors consider significant advances on the topic, in a manner that should be accessible to most readers with background in General Relativity. The work starts in section \ref{sec:sec2} with a brief motivation for the interest behind the phenomenon of superradiance in asymptotically global Anti-de Sitter spacetimes, followed by a short presentation of global AdS itself and its interesting properties. The section finishes off with an exact definition of an asymptotically AdS space, given in a few different ways. Section \ref{sec:sec3} is devoted to an introduction to the basic concepts behind superradiance illustrated by a simple example and an overview of the methods for calculating superradiant modes in different spacetimes. Chapters \ref{sec:sec4} and \ref{sec:sec5} are committed to reviewing the work done on superradiance in the specific cases of Reissner-Nordstr\"om-AdS and Kerr-AdS, respectively, with an emphasis on its effect on the stability of the two spacetimes. Finally, concluding remarks are gathered in the conclusion, followed by a list of references.
\thispagestyle{empty}
\pagenumbering{gobble}
\clearpage
\newpage
\pagenumbering{arabic}
\setcounter{page}{1}
\section{Introduction}\label{sec:sec2}
\subsection{Motivation}
Even if General Relativity was discovered just a bit more than hundred years ago, it still has not ceased to surprise us. After finding a particular solution to Einstein equation, the most tempting and logical thing to do is to investigate its behaviour under perturbations. It is in this way that one might hope to uncover the complete analytical beauty of the theory and understand more about the structure of spacetime. Moreover, there is no system in nature that is truly isolated from external influences, thus it is highly likely that the results of perturbation theory might be relevant to astrophysical observations. Following this line of thoughts, one usually starts from the simplest model there is and builds slowly on complexity. In General Relativity this corresponds to the vacuum Einstein equations with constant curvature. From the three different solutions in this case, determined by the curvature's sign, Anti-de Sitter space (with negative curvature), which is the main background spacetime in this work, stands out with a crucial difference - its conformal boundary is timelike. This implies that in order to have a well-defined Cauchy problem, one has to impose boundary conditions at infinity, with the physically relevant ones turning out to be acting like a reflecting wall. This is why AdS becomes important in the study of the other main aspect of this report - superradiance - the phenomenon in which one can extract energy from a rotating or charged black hole by scattering waves off of its horizon, depending on a certain condition satisfied by their frequency - a generalisation of the Penrose process for particles draining rotational energy from a Kerr black hole. It is Teukolsky and Press who first conjectured in \cite{press1972floating} that if the Kerr black hole is confined in a reflecting box, then the process of superradiant scattering will go on indefinitely, resulting in an exponentially growing instability. Nevertheless, black holes enclosed by perfectly reflecting walls are not something one expects to observe in nature - and even if massive fields can lead to confining potentials with trapping regions for the scattered waves - it is AdS that is the perfect system for the study of superradiance due to its reflective boundary conditions, which provide a natural confining mechanism for the radiation.\\
\hspace*{5mm}However, the importance of analysing superradiance in asymptotically AdS spacetimes does not come only due to the possibility of extending the conclusions to astrophysical systems\cite{dolan2007instability,witek2013superradiant} by juxtaposing them with the scenario of a massive field creating a confining potential around a black hole with a characteristic lengthscale similar to the radius of curvature of an AdS system. It also has implications on the stability of the spacetime - whether a solution is stable to a generic perturbation or not is vital, not only because this determines the actual significance of the theoretical construction, but also because it enables one to assess one's understanding of the phase space of the system under consideration. As it will be presented in the late part of this review - in the case of Kerr-AdS investigating its stability subject to superradiance has lead to the discovery that it is not the only stationary solution in asymptotically AdS spacetimes in four dimensions. Furthermore, there is growing evidence that its superradiant instability might have an endpoint that contradicts one of the cosmic censorship hypotheses - a result that will definitely change the way we look at General Relativity in four dimensions. On the other hand, even if not one of the main aspects of this work - the famous AdS/CFT correspondence should not be omitted. The significance of superradiance in this context comes from the fact that the effects of perturbations on the classical side can be translated into dynamical behaviour of thermal fluctuations on the field theory side. With this side remark we go back to the two points made about Kerr-AdS, as they represent some of the main results of the research in the area in recent years and them we would like to address in this essay. With this aim in mind, we take on a brief tour of the physics and mathematics behind these statements, starting from the definition of the first key ingredient in the study - pure AdS.
\subsection{Pure Anti-de-Sitter spacetime}
Anti-de-Sitter (${\rm AdS}_d$) is uniquely defined as the maximally symmetric solution of the vacuum Einstein equation with constant\footnote{With the only other two solutions with constant curvature (0 and positive) being Minkowski and De-Sitter space} negative cosmological constant $\Lambda$ in $d$ dimensions
\begin{align}
R_{ab}=\frac{2\Lambda}{d-2}g_{ab},
\end{align}
where
\begin{align}
\Lambda=-\frac{(d-1)(d-2)}{2L^2},\quad\mbox{and}\quad R_{abcd}=\frac{R}{d(d-1)}\left(g_{ac}g_{bd}-g_{ad}g_{cb}\right),\label{eq:RiemannAdS}
\end{align}
with $L$ being the radius of curvature and the characteristic lengthscale for ${\rm AdS}_d$. The second equation above implies a vanishing Weyl tensor $C_{abcd}=0$ and the symmetry group of the space is $O(d-1,2)$. The most intuitive way to visualise Anti-de-Sitter space is by embedding it in Euclidean space $\mathbb{R}^{2,d-1}$ as a hyperboloid defined by the equation
\begin{align}
X_0+X_d-\sum\limits_{i=1}^{d-1}X_i^2=L^2,\label{eq:Hyperboloid}
\end{align}
which is readily solved in coordinates $(\tau,\rho,\theta_1,...,\theta_d-3,\phi)$ by
\begin{equation*}
\begin{aligned}[c]
X_0&=L\cosh\rho\cos \tau\notag\\
X_d&=L\cosh\rho\sin \tau\notag\\
X_i&=L\sinh\rho\hat{\Omega}_i\notag\\
&\hspace*{-4.5mm}\sum\limits_{i=1}^{d-1}\hat{\Omega}_i=1\notag\\
\end{aligned}
\qquad\qquad
\begin{aligned}[c]
\hat{\Omega}_1&=\rho\cos\theta_1\notag\\
\hat{\Omega}_2&=\rho\sin\theta_1\cos\theta_2\notag\\
\vdots
\\
\hat{\Omega}_{d-2}&=\rho\sin\theta_1...\sin\theta_{d-4}\cos\theta_{d-3}\notag\\
\hat{\Omega}_{d-1}&=\rho\sin\theta_1...\sin\theta_{d-4}\sin\theta_{d-3}\notag,
\end{aligned}
\end{equation*}
where $\rho\in[0,\infty)$ and $\tau\in[0,2\pi)$, while the $\hat{\Omega}_i$'s parametrise an $S^{d-2}$ sphere with $\theta_1,...,\theta_{d-4}\in[0,\pi]$ and $\theta_{d-3}\in[0,2\pi)$\footnote{For $d=4$ one usually denotes $\theta_{d-3}$ by $\phi$}. In this way the metric for ${\rm AdS}_d$ acquires the form
\begin{align}
ds^2=L^2\left(-\cosh^2\hspace*{-0.5mm}\rho\,d\tau^2+d\rho^2+\sinh^2\hspace*{-0.5mm}\rho\, d\Omega_{d-2}^2\right),\label{eq:AdSGlobal1}
\end{align}
whereby intuitively looking at the $\rho\rightarrow0$ limit, the topology of the space can be inferred to be $S^1\times \mathbb{R}^{d-1}$, as the metric behaves like $ds^2\approx L^2\left(-d\tau^2+d\rho^2+\rho^2d\Omega_{d-2}^2\right)$.
However, due to the periodicity of $\tau$ closed timelike curves are allowed to exist in the spacetime, leading to the violation of causality. The usual approach to get around this problem (in the above way the space is also not simply connected) is to consider the universal cover of the space by effectively unrolling the circle $S^1$ and extending the limits of $\tau$ to $\tau\in(-\infty,\infty)$ (corresponding to infinitely many loops around the hyperboloid), which eliminates the possibility for closed timelike curves and changes the topology to that of $\mathbb{R}^d$. This gives the definition of biggest interest to physicists of ${\rm AdS}_d$  in global (as it covers the whole space) coordinates. There are coordinate singularities at $\rho=0$ and $\theta_i=0,\pi$, with the latter being the usual ones for spherical coordinates. Continuing in this setting, one can make the change of variables (and swapping $\tau$ for $t$)
\begin{align}
\tan\chi=\sinh\rho,\quad\mbox{with}\quad\chi\in\left(0,\frac{\pi}{2}\right),
\end{align}
leading to the metric form
\begin{align}
ds^2=\frac{L^2}{\cos^2\chi}\left(-dt^2+d\chi^2+\sin^2\chi\,d\Omega^2_{d-2}\right),
\end{align}
\begin{figure}[b]
\centering
\subfloat[\textit{Global ${\rm AdS}_d$ \eqref{eq:AdSGlobal1} (the solid-lines cylinder) which is conformally equivalent to one half ($0\leq\chi\leq\pi/2$) of the Einstein static universe (dashed cylinder).}\label{fig1a:AdScylinder}]{ %
\includegraphics[width=0.32\linewidth]{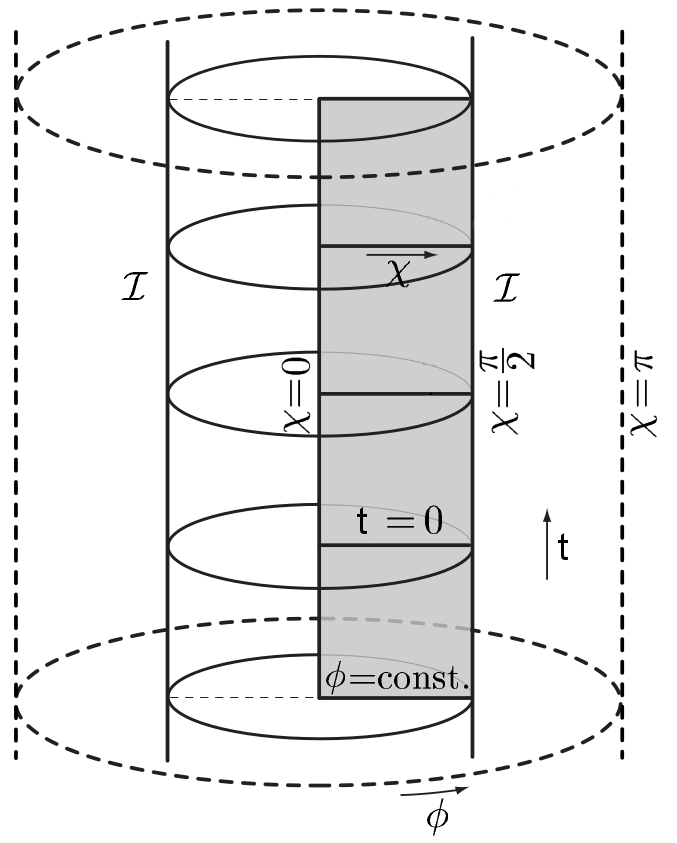}}\hfill
\subfloat[\textit{${\rm AdS}_4$ as given by \eqref{eq:AdSPenroseMetric} on the hyperboloid \eqref{eq:Hyperboloid} - covering only a part of it.}\label{fig1b:AdShyper}]{ %
\includegraphics[width=0.37\linewidth]{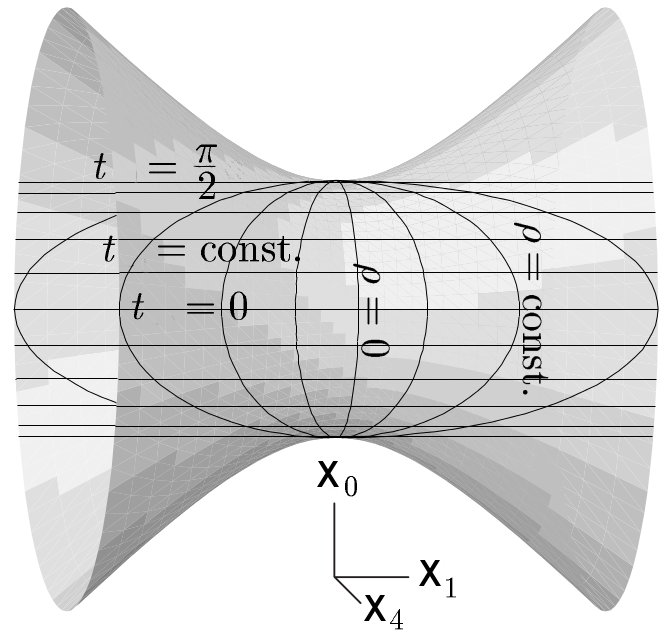}}\hfill
\subfloat[\textit{Penrose diagram of ${\rm AdS}_4$ as given by \eqref{eq:AdSPenroseMetric} after conformal compactification, whereby each point represents a 2-sphere.}\label{fig1c:AdSPenrose}]{ %
\includegraphics[width=0.25\linewidth]{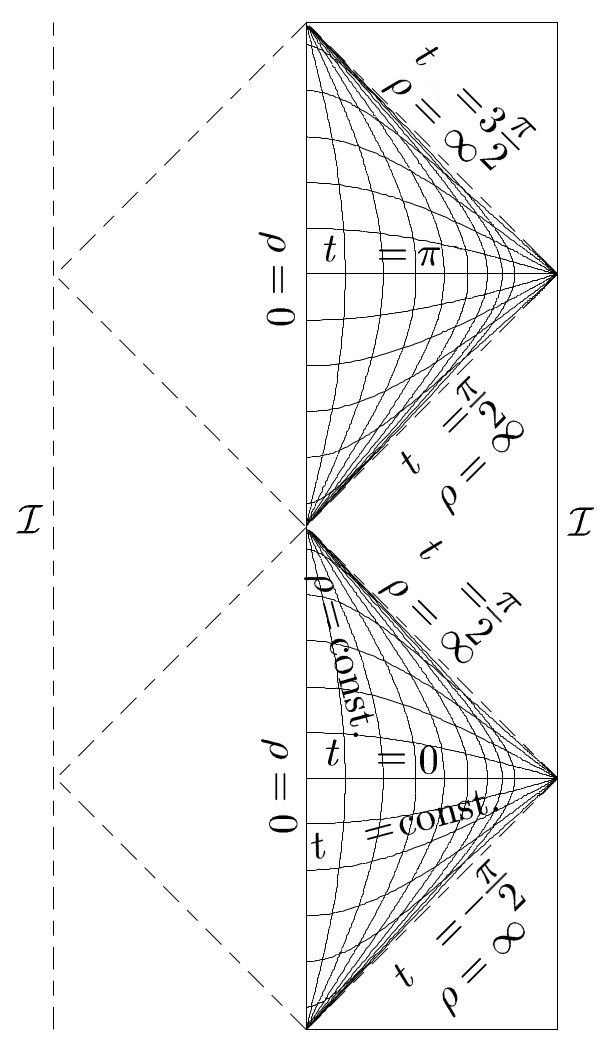}}
\caption{}
\label{fig1:AdSDiagrams}
\end{figure}
\hspace*{-2mm}which is conformally equivalent to one half of the Einstein Static Universe due to the limits of $\chi$. There are two properties of ${\rm AdS}_d$, evident from the above metric, key to for the main discussion of this work. Firstly, the conformal boundary $\mathcal{I}$ (figure \ref{fig1a:AdScylinder}), corresponding to $\chi=\pi/2$ ($\rho=\infty$), is a timelike hypersurface (in contrast to Minkowski and De-Sitter, where it is null- and spacelike, respectively), given by
\begin{align}
d\tilde{s}^2=-dt^2+d\Omega^2_{d-2},\quad\mbox{with topology }\mathbb{R}\times S^{d-2},
\end{align}
which is clearly by itself conformally flat. Its timelike character implies that Anti-de-Sitter is not globally hyperbolic - there does not exist a complete Cauchy surface in the space. Whatever family of spacelike surfaces one takes, there will always be a null geodesic that does not intersect a given such surface anywhere - e.g. surfaces of $t=const$ cover the space completely, but it is straightforward to observe that taking a null geodesic coming out from $\mathcal{I}$, at a point above the surface itself, proves the above statement in that case. This hints that to have a well-defined Cauchy problem in AdS, one must not only specify the initial data on a surface, but one must also impose appropriate boundary conditions at the conformal boundary. In fact, this was rigorously demonstrated in 1995 in \cite{friedrich1995einstein} and will be discussed in more detail in a short while. The second interesting feature of this spacetime is that null geodesics reach $\mathcal{I}$ in finite coordinate time\footnote{Whereas timelike ones never do.}, which is easily shown in another very often utilised set of coordinates for global ${\rm AdS}_d$, derived by making the following transformation in \eqref{eq:AdSGlobal1}
\begin{align}
r=L\sinh\rho,\quad\mbox{and}\quad t=L\tau
\end{align}
resulting in
\begin{equation}
ds^2=-\left(1+\frac{r^2}{L^2}\right)dt^2+\left(1+\frac{r^2}{L^2}\right)^{-1}dr^2+r^2d\Omega^2_{d-2}.\label{eq:globalAdS}
\end{equation}
By taking the normalisation condition for a radial null geodesic $g_{ab}u^au^b=0$, with $u^a=dx^a/d\tau$ - the tangent vector to the geodesic - a straightforward integration shows that
\begin{align}
\vartriangle\hspace{-1mm}t=\int\limits_{0}^{\infty}\frac{dr}{1+\frac{r^2}{L^2}}=\frac{\pi L}{2},
\end{align}
where $\vartriangle\hspace{-1mm}t$ is some finite time interval, while $r\rightarrow\infty$ corresponds to $\rho\rightarrow\infty$ where the conformal boundary $\mathcal{I}$ is located. A similar calculation for timelike geodesics leads to a divergent integral, indicating that they never reach $\mathcal{I}$.\\
\hspace*{5mm}In order to obtain the Penrose diagram of Anti-de-Sitter space, it is worth considering the $d=4$ case in yet another set of coordinates which represent a solution of \eqref{eq:Hyperboloid} - namely
\begin{align}
\begin{array}{rcl}
\left.\begin{array}{@{}l@{}}
X_0=L\sin t\vspace{1mm}\\[\jot]
X_1=L\cos t\sinh\rho\cos\theta\vspace{1mm}\\[\jot]
X_2=L\cos t\sinh\rho\sin\theta\cos\phi\vspace{1mm}\\[\jot]
X_3=L\cos t\sinh\rho\sin\theta\sin\phi\vspace{1mm}\\[\jot]
X_4=L\cos t\cosh\rho
\vspace{1mm}\end{array}\right\}\Rightarrow ds^2=L^2\left[-dt^2+\cos^2t\left(d\rho^2+\sinh^2\rho\,d\Omega^2\right)\right]\\
\end{array}\label{eq:AdSPenroseMetric}
\vspace*{7.5mm},
\end{align}
where $t\in(-\infty,\infty)$, $\rho\in[0,\infty)$, $\theta\in[0,\pi]$ and $\phi\in[0,2\pi)$ with apparent singularities at $t=\pm\frac{\pi}{2}+n\pi$, $n\in\mathbb{Z}$ and $\mathcal{I}$ is approached at $\rho\rightarrow\infty$. The above metric does not cover the whole hyperboloid, as it does not extend along the curving bits of the manifold, as seen in figure \ref{fig1b:AdShyper}, but by pulling a conformal factor of $L^2\cos^2t$ and making the transformation $t\rightarrow\tan t$ one can easily obtain an illuminating Penrose diagram for ${\rm AdS}_4$ presented in \ref{fig1c:AdSPenrose}. The worldlines of $\rho,\theta,\phi=const$ correspond to timelike geodesics (normal to surfaces of constant $t$) and as is evident from the figure - they all emanate from the same point (which without loss of generality can be taken to be $t=\pi/2$ as pure AdS is a homogeneous space) and converge at a point distance $\pi$ in $t$ away, just to defocus again and reconverge at another point further up by $\pi$ along $t$. This way it is easily observed that for a given event there are regions of space in its future light cone that cannot be reached by any timelike geodesic, including the conformal boundary $\mathcal{I}$. Therefore the infinite chain of diamonds, two of which are given in \ref{fig1c:AdSPenrose}, represents the set of points, reachable from our chosen spacetime event by timelike geodesics. This can ultimately be used as another way of seeing that a Cauchy surface in AdS cannot be found and that boundary conditions play a vital role in doing physics in AdS.\\
\hspace*{5mm}In the next section we will move on from the pure Anti-de-Sitter space and define what one means by asymptotically AdS spacetimes - which will enable us to delve later into the world of superradiance in RN-AdS and Kerr-AdS.
\subsection{Asymptotically global AdS spacetimes}\label{sec:AAdS}
The arguments in this section will be presented for $d=4$, but in general they apply for all $d\geq4$.\\
\hspace*{5mm}As mentioned in the previous section, AdS is the maximally symmetric solution of the vacuum Einstein equation with negative cosmological constant, henceforth, for $\Lambda<0$ it plays the role that Minkowski plays for flat spacetimes. It is therefore not only natural to think about the concept of being asymptotically AdS (by which in this work we mean exclusively asymptotically global AdS), but it is also needed when one wants to explore more thoroughly the properties and dynamics of black holes and matter in Anti-de Sitter. This is most often carried out with the tools of perturbation theory which implies that one should find a proper way of introducing perturbations, such that they are generalised enough in order to reveal new things about the system, while keeping the spacetime well-defined and preserving its structure - by which, as in the case of asymptotically flat spacetimes, it is understood the asymptotic one. Therefore, a definition of an asymptotically AdS spacetime is required and it is supplied by \cite{henneaux1985asymptotically} in the form of three requirements on the imposed boundary conditions at spatial infinity:
\begin{enumerate}[label=\protect\color{black}$\blacklozenge$]
\setlength\itemsep{-1mm}
\item They should be invariant under the global AdS symmetry group $O(3,2)$
\item They should make the surface integrals associated with the generators of the AdS group $O(3,2)$ finite
\item They should include the asymptotic behaviour of the Kerr-AdS metric
\end{enumerate}
The first of these is straightforward - if it were not the case, then a symmetry transformation could take an allowed set of conditions to one which is not, making the whole procedure meaningless. The second requirement is based on the canonical formulation of the problem - wherein one rewrites the otherwise vanishing Hamiltonian of the theory by adding surface integrals corresponding to the generators of the $O(3,2)$ group, which make the variational derivatives of the canonical variables well defined, thus enabling the exploration of the dynamics of the system. If these surface terms are not finite, the newly written Hamiltonian will diverge, hence the second condition. The last requirement is what ensures that the boundary conditions are not too restrictive in the sense that they allow for metrics that are of interests to physicists to be considered and Kerr-AdS, as in the case of $\Lambda=0$ and pure Kerr, is what is reasonably expected to be the configuration to which isolated systems asymptote in AdS settings. By considering possible perturbations that obey these three points (most simply achieved for gravitational ones by acting with $O(3,2)$ on the metric of Kerr-AdS itself, as it has been defined to be asymptotically AdS, and looking at the decays of the components at spatial infinity), it can be shown that $O(3,2)$ will be realised as the asymptotic symmetry group at spatial infinity and given that `reflective' boundary conditions are imposed there, then $\mathcal{I}$ will be conformally flat. The last bit can be ensured by looking at the Weyl tensor and its asymptotic behaviour, but involves some technicalities and will not be presented here - a more detailed discussions can be found in \cite{henneaux1985asymptotically,ashtekar2000asymptotically} and references therein - crucially, certain requirements on the decay of the components of the Weyl tensor are derived. On a further note, reflective in this context implies that allowed perturbations of Anti-de-Sitter should be described as standing waves with a node at the conformal boundary $\mathcal{I}$. Combining these observations with the second condition above implies that the asymptotic structure of AdS is conserved - that is the boundary metric is preserved.\\
\hspace*{5mm}The notion of an asymptotically AdS space can also be formulated in the spirit of the textbook definition of asymptotic flatness based on conformal compactification \cite{ashtekar2000asymptotically}, with the obvious difference that spatial infinity is required to approach that of AdS, rather than Minkowski - i.e. have an $\mathbb{R}\times S^2$ topology and vanishing fluxes across it. However, this will not be laid out here and the interested reader is referred to the above article for a good presentation of the topic.\\
\hspace*{5mm}Finally, a more physically intuitive elucidation of being asymptotically AdS, in light of our definition of Anti-de-Sitter \eqref{eq:RiemannAdS}, and motivated by conformal compactification comes from \cite{skenderis2002lecture} - namely: \textit{`An Asymptotically AdS metric is a conformally compact Einstein metric'}. This is actually fairly straightforward to deduce - taking a conformally compact manifold\footnote{For a discussion on the definition of a conformally compact manifold the reader is redirected to \cite{penrose1988spinors}.} $M$ with metric $g$, such that
\begin{align}
\tilde{g}=z^2g,\label{eq:ConfMetric}
\end{align}
where $\tilde{g}$ is the conformal metric and $z$ a smooth positive function on $M$, and working to leading order in $z$ for $z\rightarrow0$ (where the conformal boundary is located), by just plugging \eqref{eq:ConfMetric} into the definition of the Christoffel symbols and then taking the leading order contribution and inserting it in the definition of $R_{\mu\nu\rho\sigma}$ in terms of $\Gamma^\lambda_{\mu\nu}$, it is found that
\begin{align}
R_{\mu\nu\rho\sigma}(g)=\tilde{g}^{\tau\lambda}\partial_{\tau}z\partial_{\lambda}z\left(g_{\mu\rho}g_{\nu\sigma}-g_{\mu\sigma}g_{\nu\rho}\right)+\mathcal{O}\left(z^{-3}\right),\label{eq:RiemannConformal}
\end{align}
Furthermore, by demanding that $g$ is a solution to the vacuum Einstein equation with negative cosmological constant, it can be derived that $\tilde{g}^{\tau\lambda}\partial_{\tau}z\partial_{\lambda}z=1/L^{2}$ and thus one sees in the limit $z\rightarrow0$ that equation \eqref{eq:RiemannConformal} approaches \eqref{eq:RiemannAdS}, hence the above definition.\\
\hspace*{5mm}
Having given a brief introduction to Anti-de-Sitter space, in the next section the main topic of this work - superradiance - will be introduced, but firstly in a heuristic fashion by using an example in the flat case of Kerr, while later its natural extension to black holes in AdS will be considered.
\section{Superradiance}\label{sec:sec3}
\subsection{Simple example and discussion}\label{sec:SimpleEx}
The idea of superradiance or superradiant scattering is usually introduced as a generalisation of the Penrose process to waves, but this will not be the approach taken here\footnote{For a good description of the Penrose process the reader is referred to \cite{wald2010general}.}. It will be rather illustrated with a simple example in the Kerr spacetime, followed by a short discussion of its appearance in different contexts and its implications on the studied systems.\\
\hspace*{5mm}Consider the Kerr spacetime with two spacelike surfaces $\Sigma$ and $\Sigma'$, both stretching from $i^0$ to $\mathcal{H^+}$, with $\Sigma'$ being entirely to the future of $\Sigma$. Furthermore, define $H$ and $H'$ as the intersections of $\Sigma $ and $\Sigma'$ with the future event horizon and take $\mathcal{N}$ to be the part of $\mathcal{H}^+$ from $H$ to $H'$. Moreover, as Kerr is a stationary spacetime, total energy of matter on a spacelike hypersurface can be defined naturally as
\begin{align}
E(\Sigma)=-\int_{\Sigma}\star J,
\end{align}
where $J_a=-T_{ab}k^b$ is the conserved energy-momentum 4-vector and $k^b$ is the timelike Killing vector field. Using this, it is easily shown that
\begin{align}
E(\Sigma')-E(\Sigma)=\int_{\mathcal{N}}\star J,\label{eq:HorizonFlux}
\end{align}
which gives a definition of the flux across the horizon as the difference between the energies of the two spacelike hypersurfaces. To continue, take matter to be given by a massless scalar field with stress-energy tensor $T_{ab}=\partial_a\psi\partial_b\psi-\frac{1}{2}g_{ab}\partial^c\psi\partial_c\psi$. As the spacetime is stationary and axisymmetric with corresponding commuting Killing vector fields - $\partial_t$ and $\partial_\phi$ - the scalar field can be decomposed as $\psi(t,r,\theta,\phi)=Re\left[\psi_0\left(r,\theta\right)e^{-i\omega t}e^{im\phi}\right]$, where $\omega$ is a frequency and $m$ - an integer - the azimuthal quantum number. By a simple brute force calculation it can be quickly shown that for $0<\omega<m\Omega_H$ the right hand side of equation \eqref{eq:HorizonFlux} is positive. To this end, take the 4D Kerr metric in Kerr coordinates $(v,r,\theta,\chi)$ and note that $\mathcal{N}$ is a three-dimensional manifold, hence in order to carry out the integration of the three form $\star J$ one just needs to specify
\begin{equation}
\left(\star J_{\nu\theta\chi}\right)_{r=r_+}=\sqrt{-\det g}\,\epsilon_{\nu\theta\chi\mu}J^\mu=\sqrt{-\det g}\,g^{r\mu}J_\mu,
\end{equation}
where the determinant and the inverse metric can be found by straightforward computations or using \textit{Mathematica} and look as $\det g=-\Sigma^2\sin^2\theta$ and $g^{r\mu}\partial_\mu=\frac{1}{\Sigma}\left[\Delta\partial_r+(r^2+a^2)\left(\partial_v+\Omega_H\partial_\chi\right)\right]$, with $\Sigma=r^2+a^2\cos^2\theta$, $\Omega_H=\frac{a}{r^2+a^2}$, $\Delta=r^2-2Mr+a^2=(r-r_+)(r-r_r)$ and $a$ the rotation parameter. Using the fact that $J_\mu=g(J,\partial_\mu)=\langle J,\partial_\mu\rangle$ and the definition of the horizon generating Killing vector $\xi^a=k^a+\Omega_Hm^a=\left((\partial_v)^a+\Omega_h(\partial_\chi)^a\right)$ one obtains the required quantity
\begin{equation}
\left(\star J_{\nu\theta\chi}\right)_{r=r_+}=\sin\theta(r_+^2+a^2)\xi^aJ_a.
\end{equation}
Henceforth, to determine the sign of \eqref{eq:HorizonFlux} one just needs to look at $\xi^aJ_a=-\xi^aT_{ab}k^b=-\xi^a\partial_a\psi\partial_b\psi k^b$, where the second term that would come from the given stress-energy tensor vanishes due to $\xi\cdot k=0$ on $\mathcal{H^+}$, as a consequence of the horizon invariance under the isometries of the spacetime, implying that KVFs should be tangent to it and thus orthogonal to its generators (the horizon is Killing). Finally, a little bit of differentiation of the given scalar field leads to the final answer, which takes the form
\begin{equation}
\omega\left(m\Omega_H-\omega\right)\geq0,
\end{equation}
giving the condition $0<\omega<m\Omega_H$ for \eqref{eq:HorizonFlux} to be positive. This simple results has the remarkable implication that energy can be extracted from the black hole by scattering waves off of it - the phenomenon of superradiance. Now, it is only natural for a theoretical physicists to try to enclose the superradiant object in question with a reflecting surface, so that the waves can go on scattering back and forth indefinitely, thus draining all the energy of the black hole. This can be achieved by surrounding the object with a giant mirror for example - which was first proposed in \cite{zel1972amplification} for the case of electromagnetic waves impinging upon a conducting rotating cylinder. A more realistic pathway towards achieving superradiance in Kerr, which has also recently started to attract more attention in the astrophysics community\footnote{For an interesting read on the topic the following two papers are recommended \cite{dolan2007instability,witek2013superradiant}}, is by considering a massive scalar field instead of a massless one. The addition of the mass term leads to a potential in the Klein-Gordon equation for the field that exhibits a local minimum between the event horizon of the black hole and spatial infinity, wherein scattered waves can get `trapped' and reflected backwards, so as to be amplified again due to superradiance \cite{cardoso2005superradiant}. By looking at the asymptotic behaviour of the potential one sees that this is always the case in $d=4$, as long as $\mu<\omega$, where $\mu$ is the mass of the scalar field. Of course, there is a separate condition on the frequency $\omega$ itself for when the wave modes are superradiant that depends on the rotation parameter and the radius of the black hole (for fixed $\mu$ and $m$).\\
Naturally, as the reader might have already guessed there is an obvious candidate to investigate superradiance in and this is asymptotically AdS spaces, due to the timelike nature of spatial infinity and the reflective boundary conditions there. These imply that Anti-de Sitter acts just like a confining box and any waves (moving at the speed of light) scattered from the bulk outwards will eventually reach spatial infinity (as eluded to earlier) and get reflected backwards there. Depending on the type of perturbations (scalar, electromagnetic or gravitational) and the spacetime under consideration, the situation may be quite different. For Schwarzschild-AdS\cite{cardoso2001quasinormal,michalogiorgakis2007low,dias2013boundary} it has been shown that superradiance does not occur\footnote{Which can be expected once one has looked in more detail into the requirements for the appearance of superradiance - which will be done later in the section.}, while for RN-AdS and Kerr-AdS there are both quasinormal(QNM) and superradiant modes present. The former are usually defined in a physics context as wave solutions which are purely outgoing at spatial infinity $\psi\approx e^{-i\omega(t-r^*)}$ and solely ingoing near the horizon $\psi\approx e^{-i\omega(t+r^*)}$, where $r^*$ is the usual tortoise coordinate ($dr^*=dr/f(r)$ for Schwarzschild-AdS) and $\omega\in\mathbb{C}$. Mathematically, QNMs appear when the two solutions of the wave equation under consideration are linearly dependent, with the coefficient of proportionality being the complex QNM frequency. It is interesting to note\cite{kokkotas1999quasi} that an analysis of the behaviour of the QNM eigenfunctions shows that their decay in time depends on the asymptotic properties of the potential and for it to be exponential in nature the potential has to be vanishing outside a certain region centred at the origin. Therefore, the usual identification of QNMs with exponentially decaying perturbations might be a bit naive. Nevertheless, for the spacetimes investigated in this work the potential is always asymptotically vanishing and thus the QNMs at large distances from the origin will be dying off with time. On the other hand, the superradiant modes which we introduced as growing in time and which are the main focus of this work, seem to cause the RN- and Kerr-AdS systems to take on two at first similar paths that later split in opposite directions.\\
\hspace*{5mm}It is the stability of the spacetimes under investigation that is being referred to at the end of the last paragraph and it is one of the main reasons why superradiant scattering is interesting. Stability is important from the point of view that an instability with a timescale that is not comparable with the age of the universe will most probably lead to a very small number of representatives of the system in question in Nature. Moreover, superradiance in an astrophysical context might lead to observable gravitational wave emissions and may be used to constrain certain beyond-the-standard-model-physics models\cite{arvanitaki2011exploring,pani2012black}. Also, back to AdS and relating to the AdS/CFT correspondence - perturbations of black holes in the bulk are linked to ones on the boundary CFT, thus the time evolution of the quasinormal and superradiant modes is dual to the evolution of fluctuations of the field theory.\\
\hspace*{5mm}Elucidating more on the instabilities - the defining prerequisite for their existence in a spacetime is the presence of growing in time perturbations - that is, superradiant modes - and the exciting consequences thereof are that they might lead to the transition of the system to a different state, the formation of new objects or redistribution of energy between the excited modes of the perturbation - all of which are interesting possibilities that might uncover some new black hole physics, which is why, in recent years, a lot of effort has been put in understanding the effect of superradiant scattering in asymptotically AdS spacetimes.\\
\hspace*{5mm}As promised earlier, a brief discussion on the requirements for the presence of superradiant modes will be now presented\cite{hawking1999charged}. Take the first law of black hole mechanics for a rotating black hole
\begin{align}
dM=\frac{\kappa}{8\pi}dA+\Omega_HdJ,
\end{align}
which relates the change in mass and angular momentum, due to a linearised perturbation, to the change in the horizon area. Then, consider a scalar field (the same argument can be generalised to any electromagnetic or gravitational perturbation in a straightforward way) with a stress energy tensor given as before by
\begin{align}
T_{ab}=\partial_a\psi\partial_b\psi-\frac{1}{2}g_{ab}\partial^c\psi\partial_c\psi.
\end{align}
Looking at the $\tensor{T}{^r_t}$ and $\tensor{T}{^r_\phi}$ components, corresponding to the net radial flux of energy and angular momentum, respectively, it is easy to show that the ratio of mass to angular momentum carried in the black hole by the wave results in
\begin{align}
\frac{dM}{dJ}=\frac{\omega}{m},
\end{align}
where as before the scalar field has been decomposed according to the isometries of the spacetime - $\psi(t,r,\theta,\phi)=\psi_0\left(r,\theta\right)e^{-i\omega t}e^{im\phi}$. Consequently, referring to the second law of black hole mechanics, which informs us that classically
\begin{align}
dA\geq0,
\end{align}
for a field scattering off the black hole, given that it obeys the dominant energy condition, it is straightforward to derive that energy can be extracted from the black hole under the condition that
\begin{align}
\omega<m\Omega_H.
\end{align}
Reasoning along exactly the same lines for a charged black hole, where the angular momentum is replaced by the electrostatic potential of the black hole, leads to the analogous conclusion that superradiance appears given that
\begin{align}
\omega<q\Phi,
\end{align}
where $q$ is the charge of the scalar field and $\Phi$ the electrostatic potential difference between the horizon and infinity. This can also be written as $\omega<q\frac{Q}{r_+}$, treating the extracting superradiant mode as being just outside the black hole, thus permitting the usual approximation for a homogeneous spherically symmetric lump of charge ($r_+$ is the radius of the black hole).\\
\hspace*{5mm}The above conditions do not actually imply that a given rotating or charged black hole will suffer from superradiant instabilities. In order to show this, one has to go through the linearised Einstein equations for a given perturbation and show that modes with the desired frequencies actually exist, which is definitely not a straightforward process. However, in the case of asymptotically AdS spacetimes it has very recently\cite{green2015superradiant} been proven mathematically that any asymptotically AdS black hole with a Killing horizon, whose corresponding Killing field becomes spacelike in some region of space\footnote{Which defines it as an ergoregion.}, will be linearly unstable due to superradiance of gravitational perturbations\footnote{Technically, what is shown in the paper is that the system does not go back to equilibrium, which does not rule out the case of oscillations around it with a constant amplitude. Even though this will clearly not lead to an instability, it is a special case that usually can be ruled out with befitting confidence by numerical results.}. This result is not going to be rederived here, but one of its immediate implications is that Kerr-AdS is linearly unstable to superradiance, which will be covered in great detail later in this work. It should be noted that the above theorem can not be straightforwardly implied to charged black holes in AdS, even if a notion of a generalised ergoregion can be introduced for them - a little discussions about this with a reference is given at the end of \cite{green2015superradiant}.\\
\hspace*{5mm}Having presented the phenomenon of superradiance in a short manner, we are going to move on and illustrate in the next section some of the popular methods for actually calculating the superradiant modes of different black hole spacetimes subject to various perturbations. This is also a good place to refer the reader to a long review on the subject that is extremely helpful in obtaining references - \cite{brito2015superradiance}.
\subsection{Calculating quasinormal and superradiant modes\\in asymptotically AdS spacetimes}
It should be made clear that from this point onwards only asymptotically AdS spacetimes in 4D are investigated - in particular Reissner-Nordstr\"{o}m-AdS (RN-AdS) and Kerr-AdS, with an occasional reference to Schwarzschild-AdS.
\subsubsection{Scalar fields and wave equations}\label{subsec:Scalar}
Generally, perturbations are devised in three types - scalar, electromagnetic and gravitational, and in this section it will be the simplest type of perturbations - scalar fields - that will be considered first\cite{cardoso2004small,uchikata2009scalar,uchikata2011quasinormal,bosch2016nonlinear}. The reason being that in this way the reader will be gently introduced to the logical flow behind this type of calculations, which are fairly similar in character, even if they differ quite a lot in the complexities of their specifics. A further simplification in this case, for asymptotically AdS spacetimes, comes from the fact that it is sufficient to consider massless fields due to the reflective nature of the boundary at infinity. Therefore, one starts from the Klein-Gordon equation for the scalar field
\vspace*{4.5mm}
\begin{align}
\begin{array}{rcl}
\nabla_\mu\nabla^\mu\Phi=
\smash{\left\{\begin{array}{@{}l@{}}
\frac{1}{\sqrt{-g}}\partial_\mu\left(\sqrt{-g}g^{\mu\nu}\partial_\nu\Phi\right)=0,\quad\hfill\mbox{for Kerr-AdS}\vspace{3mm}\\[\jot]
\left(\nabla_\mu-ieA_\mu\right)\left(\nabla^\mu-ieA^\mu\right)\Phi=0,\quad\mbox{for RN-AdS}\vspace{1mm}\\[\jot]
\end{array}\right.}\\\vspace*{0.5mm}
\end{array},\label{eq:waveEq}
\end{align}
where $\Phi$ is the scalar field and $A_\mu$ is the Maxwell gauge field. Afterwards, a separation ansatz can be imposed
\begin{align}
\Phi(t,r,\theta,\phi)=e^{-i\omega t}e^{im\phi}R(r)S(\theta),\label{eq:ScalarField}
\end{align}
where as before the field has been decomposed into Fourier modes taking advantage of the isometries of the background, corresponding to the Killing vectors $\partial_t$ and $\partial_\phi$, with $\omega$ a complex frequency and $m$ an integer. For RN-AdS, due to spherical symmetry, the $e^{im\phi}$ and $S(\theta)$ parts can be combined into the usual spherical harmonics $Y_{lm}(\theta,\phi)$ with $l$ and $m$ the usual angular momentum and azimuthal quantum numbers. By plugging the decomposition \eqref{eq:ScalarField} back into \eqref{eq:waveEq} two equations are obtained - a radial and an angular one, whereby the separation constant (which can be shown to be the same for both equations) corresponds to the eigenvalue of the angular equation. Continuing analytically at this point is usually done by defining a near-horizon region $r-r_+\ll1/\omega$, where $r_+$ is the location\footnote{It is the largest root of $\Delta$ in the usual notation for the metrics of RN-AdS and Kerr-AdS.} of the future event horizon $\mathcal{H^+}$, and a far region $r-r_+\gg r_+$. In the former the contribution of the Cosmological constant can be neglected and $r\approx r_+$, which leads to a number of simplifications of the radial equation, which after a suitable transformation can be turned into a standard hypergeometric differential equation, whose general solution is readily available. Of course, it should not be forgotten that the relevant boundary conditions have to be imposed and near the horizon this implies that only ingoing waves are allowed, as one does not expect perturbations to be coming out of the black hole (in the classical picture). Deducing which coefficient should be set to zero in the general solution in terms of hypergeometric functions can be done by performing a Frobenius analysis of the radial equation around the horizon\footnote{That is - expanding $R(r)$ in the appropriate power series of the form $r^k\sum\limits_{n=0}^{\infty}c_nr^n$, where $c_n$ are some coefficients.}. Afterwards, turning to the far-region - there the effects of the black hole can be neglected and the radial equation reduces to that of pure AdS with the subtle difference that the inner boundary in that case is at $r_+$ and not at $r=0$. Nevertheless, progress is achieved in the same way as in the near-horizon region and after a suitable substitution, the equation reduces to a standard hypergeometric equation. Then again the relevant boundary conditions have to be taken into account, however, in this situation more caution is required. As it was already mentioned in the section on asymptotically AdS spacetimes, at spatial infinity perturbations have to behave like a standing wave with a node there, meaning that in general both the ingoing and outgoing waves have to be considered. Nevertheless, after undertaking a Frobenius analysis near $\mathcal{I}$ it is seen that for a scalar field this is straightforward as, in order to avoid the field diverging, one of the coefficients has to be set to zero and the surviving part of the solution meets the requirements on the decay of the Weyl tensor from \cite{henneaux1985asymptotically}. Although at present completing the solution analytically for the whole phase space of black holes is not possible, a restriction to $r\omega\ll1$ provides a way out\footnote{Plus further taking $\frac{a}{r_+}\ll1$ for Kerr-AdS}. In this regime the near- and far-regions overlap in the zone $r_+\ll r-r_+\ll1/\omega$ and it can be shown that the condition $r_+\omega\ll1$ is equivalent to working in the regime $\frac{r_+}{L}\ll1$  - i.e. small black holes. This equivalence will be derived here for Kerr-AdS, but it is analogous and simpler in the case of RN-AdS. So starting from $\frac{r_+}{L}\ll 1$, taking the condition for extremality for small radii
\begin{align}
a\leq r_+\sqrt{\frac{3r_+^2+L^2}{L^2-r_+^2}},\quad\mbox{for }r_+<\sqrt{3}L,\label{eq:aExtremal}
\end{align}
and expanding in series for small $r_+$ to get $a\leq r_++\mathcal{O}(r_+^3)$, it is immediately obvious that $\frac{a}{L}\ll1$. Afterwards, arguing that $\frac{r_+}{L}\ll 1$ means that the real part of the frequencies would be of the order of those in pure AdS which are calculated in \cite{dias2013boundary} and behave as $\omega L\approx\mathcal{O}(1)$, it is easily observed that one also gets $r_+\omega\ll1$ and $a\omega\ll1$ (which are also used in the simplification of the radial equation and are needed for the Kerr-AdS condition $a/r_+\ll1$). This argument is sensible, as for a tiny black hole, one would generally expect the effect on the spacetime to be fairly negligible throughout most of it. Therefore, for small black holes the near-horizon and far-region solutions can be matched asymptotically in the intermediate, overlapping zone. This requires deriving the asymptotic behaviour of the former for large $r$ and of the latter for small, which can be straightforwardly achieved by using the properties of hypergeometric functions. The result of this procedure is a quantised spectrum for the frequency $\omega$, whereby it turns out that the sign of the imaginary part depends on a condition on the real part - that is
\begin{align}
&\mbox{Re}(\omega)-q\frac{Q}{r_+}<0\quad\Rightarrow\quad\mbox{Im}(\omega)>0,\quad\mbox{for RN-AdS}\label{eq:ineqRN}\\
&\mbox{Re}(\omega)-m\Omega_H<0\quad\Rightarrow\quad\mbox{Im}(\omega)>0,\quad\mbox{for Kerr-AdS},\label{eq:ineqKerr}
\end{align}
where for an equality in the conditions for the real part of the frequencies (that is - Re$(\omega)=q\frac{Q}{r_+}$ or Re$(\omega)=m\Omega_H$) the imaginary part vanishes, while the reversed inequalities expectedly lead to Im$(\omega)<0$. The sign of the imaginary part is a clear indicator for the nature of the mode under consideration, as evident from equation \eqref{eq:ScalarField} - for Im$(\omega)<0$ the wavefunction is exponentially decaying in time - thus the mode is damped and it is identified as a QNM. For Im$(\omega)>0$ the scalar field perturbation has an exponential time growth and thus corresponds to a superradiant mode as it bounces back and forth between the horizon and the conformal boundary. It is important to note that the above relations can be obtained without actually deriving the frequency spectrum. This can be achieved in a similar way to the example with the real scalar field at the beginning of the previous section. Define, as before, the total flux through a hypersurface as
\begin{align}
\mathcal{F}(\Sigma)=-\int_{\Sigma}\star J,
\end{align}
where $J_a=-\mathcal{T}_{ab}\xi^b$ is the conserved 4-current associated with a given Killing vector field $\xi^b$, which for RN-AdS and Kerr-AdS can be either $\partial_t$ or $\partial_\phi$ in which case $J$ represents the energy or angular momentum 4-vector, respectively, and $F(\Sigma)$ - the energy or angular momentum flux through $\Sigma$. $\mathcal{T}_{ab}$ is the stress-energy tensor of the perturbation, which from the linearised Einstein equation can be shown to be proportional to the Landau-Lifschitz pseudotensor\cite{landau1971classical}, which most importantly is expressible only in terms of metric components - which in linearised theory are the ones of the perturbation. Of course, for RN-AdS there is no notion of angular momentum (hence the above expression will always vanish for $\partial_\phi$) and one instead defines the electric charge on a hypersurface in an analogous way
\begin{align}
\mathcal{Q}(\Sigma)=-\int_{\Sigma}\star j,
\end{align}
where $d\star F=-4\pi\star j$ - with $F$ the usual Maxwell field-strength tensor. Computing these integrals is a rather long and not very exciting task\footnote{For Kerr-AdS one can take directly $\xi=\partial_t+\Omega_H\partial_\phi$, which is both the horizon generator and the normal to it.}, usually done in ingoing Eddington-Finkelstein cooridnates, but the final answers reduce to the inequalities presented above - which is a good indication that the perturbative expansions in the two regions and the asymptotic matching in the overlapping zone provide a good approximation in the regime of small black holes. Confirming this analysis and exploring the rest of the phase space for black holes is then done numerically - \cite{cardoso2004small,cardoso2006classical,uchikata2009scalar,cardoso2014holographic,uchikata2011quasinormal,dias2012hairy,bosch2016nonlinear}.
\subsubsection{The Newman-Penrose and Teukolsky's formalisms}\label{subsec:NPFormalism}
A different approach to the above calculations, which nevertheless in the case of a scalar field perturbation reduces to what is laid out above, but is applicable to all types of perturbations, is given by the so called Teukosly's formalism\cite{teukolsky1973perturbations,chandrasekhar1998mathematical,kramer1980exact}, which is based on the Newman-Penrose (NP) tetrad formalism, which will be given a brief introduction here, but is very well presented in \cite{chandrasekhar1998mathematical} and in almost any other textbook on General Relativity. Technically, Teukolsky's approach was initially devised for rotating black holes, but it applies equally well to the Schwarzschild spacetime in the limit of a vanishing rotation parameter $a\rightarrow0$ and can then straightforwardly be generalised to RN. To begin with the basics - a tetrad formalism uses a tetrad basis - four linearly independent vector fields $\tensor{e}{_{(a)}^i}$, $a=\{1,2,3,4\}$ - which set up at each point of the spacetime a basis of four vectors that satisfy
\begin{align}
\tensor{e}{_{(a)}^i}\tensor{e}{^{(b)}_i}=\tensor{\delta}{_{(a)}^{(b)}},\quad\mbox{and}\quad \tensor{e}{_{(a)}^i}\tensor{e}{^{(a)}_j}=\tensor{\delta}{^i_j},\quad\mbox{where }\tensor{e}{_{(a)i}}=\tensor{g}{_{ij}}\tensor{e}{_{(a)}^j},
\end{align}
and $\tensor{e}{^{(b)}_i}$ is the matrix inverse of $\tensor{e}{_{(a)}^i}$. The bracketed indices indicate the tetrad components, whereas the ones without a bracket are the usual tensor indices. Also part of the definition is
\begin{align}
\tensor{e}{_{(a)}^i}\tensor{e}{_{(b)}_i}=\tensor{\eta}{_{(a)(b)}},
\end{align}
with $\tensor{\eta}{_{(a)(b)}}$ a constant symmetric matrix, which is used to lower and raise the tetrad indices. The general idea behind the adoption of a tetrad basis is that with an appropriate choice it should be possible to get a better handle of the underlying symmetries of the system under consideration. Of course, this implies that choosing the tetrad vectors is not a trivial process. With the above definitions it is a simple exercise to show that
\begin{align}
\tensor{e}{_{(a)i}}\tensor{e}{^{(a)}_j}=\tensor{g}{_{ij}}.\label{eq:MetricCompsTetrad}
\end{align}
The idea of the tetrad formalism is to project all the quantities of interest onto it and solve the relevant equations for them in this basis, whereby the projections are defined as
\begin{align}
\tensor{T}{_{(a)(b)}}=\tensor{e}{_{(a)}^i}\tensor{e}{_{(b)}^j}\tensor{T}{_{ij}}=\tensor{e}{_{(a)}^i}\tensor{T}{_{i(b)}},\\
\tensor{T}{_{ij}}=\tensor{e}{^{(a)}_i}\tensor{e}{^{(b)}_j}\tensor{T}{_{(a)(b)}}=\tensor{e}{^{(a)}_i}\tensor{T}{_{(a)j}},
\end{align}
with the obvious generalisation for tensors of any rank. From here onwards brackets around indices will be omitted - adopting the convention that earlier letters in the Latin alphabet correspond to tetrad components, while the later ones designate tensor indices. Furthermore, by choosing a coordinate basis, the tetrad can be written as linear combinations of tangent vectors
\begin{align}
\tensor{e}{_{a}}=\tensor{e}{_{a}^i}\partial_i,
\end{align}
which identifies them as directional (with respect to the tetrad basis vectors) derivatives and additionally implies that differentiating with respect to the tetrad indices can be expressed in terms of the usual partial and covariant derivatives of tensor quantities
\begin{align}
\tensor{A}{_{a,b}}=\tensor{e}{_{a}^j}\tensor{e}{_{b}^i}\nabla_iA_j+\gamma_{cab}A^c,\label{eq:TetradDerv}
\end{align}
where $\gamma_{cab}$ are called the Ricci-rotation coefficients, defined by
\begin{align}
\gamma_{cab}=\tensor{e}{_c^k}\tensor{e}{_b^i}\nabla_i\tensor{e}{_{ak}}\quad\mbox{and}\quad\gamma_{cab}=-\gamma_{acb}, 
\end{align}
and are the second key ingredient, after the tetrad basis vectors, of a given tetrad formalism as will become clearer in a bit, when the Petrov classification of spacetimes is reviewed. The rotation coefficients can be viewed alternatively as a connection in this basis, as is easily identifiable from the following definition
\begin{align}
\nabla_i\tensor{e}{_{ak}}=\tensor{e}{^c_k}\gamma_{cab}\tensor{e}{^b_i}\quad\Rightarrow\quad\nabla_i\tensor{e}{_a^k}=\tensor{\gamma}{_a^k_i},
\end{align}
which makes it possible to rewrite equation \eqref{eq:TetradDerv} as
\begin{align}
\tensor{e}{_{a}^j}\tensor{e}{_{b}^i}\nabla_iA_j=\tensor{A}{_{a,b}}-\eta^{cd}\gamma_{cab}A_d=A_{a|b}
\end{align}
where the RHS of the equation has been identified with the \textit{intrinsic derivative} of $A_a$ in the direction $e_b$. This quantity will not be explicitly needed here, but it is an essential part of the tetrad formalism and is used in many derivations of interest, hence its mentioning. The above definitions provide all the necessary tools to project all the relevant quantities - like the Riemann, Weyl and Ricci tensors onto the tetrad basis and obtain the Ricci- and Bianchi-identities in terms of tetrad components. These will not be presented here, as the expressions are quite space-consuming, but they can be readily found in many textbooks - \cite{chandrasekhar1998mathematical} with the mostly negative convention or \cite{kramer1980exact} for the predominantly positive one. Lastly, before formally introducing the Newman-Penrose choice of tetrad basis, we will mention that quite often\footnote{Especially in simpler calculations and in university courses on General Relativity} the natural choice for a tetrad basis is an orthonormal one in which case $\eta_{ab}$ takes the form of the Minkowski metric. For example, for asymptotically flat Schwarzschild one can take $e^1=(1-2M/r)^{1/2}dt$, $e^2=(1-2M/r)^{-1/2}dr$, $e^3=rd\theta$ and $e^4=r\sin\theta d\phi$, where it is sometimes easier to define the tetrad basis in terms of covectors.\\
\hspace*{5mm}The Newman-Penrose formalism consists in a special choice of the tetrad basis vectors, based on the belief of Roger Penrose that the causal structure of a spacetime is one of its key elements, which is also evident from the Penrose diagrams he introduced. Therefore, unsurprisingly, the NP tetrad basis consists of four null vectors: $\textbf{l}, \textbf{n}, \textbf{m}$ and $\bar{\textbf{m}}$, where the former two are real, while the latter are complex conjugates of each other. They satisfy
\begin{gather}
\textbf{l}\cdot\textbf{m}=\textbf{l}\cdot\bar{\textbf{m}}=\textbf{n}\cdot\textbf{m}=\textbf{n}\cdot\bar{\textbf{m}}=0\\
\textbf{l}\cdot\textbf{n}=-1\quad\mbox{and}\quad\textbf{m}\cdot\bar{\textbf{m}}=1
\end{gather}
where the latter two relations are not strictly necessary, but in most cases simplify computations significantly as one does not need to worry about various coefficients arising while raising and lowering indices and playing with directional and intrinsic derivatives in tensor notation. In this formalism both the directional derivatives and the rotation coefficients, which are called spin coefficients now, are given special symbols
\begin{gather}
D=\textbf{l}^k\partial_k,\quad \tilde{\Delta}=\textbf{n}^k\partial_k,\quad \delta=\textbf{m}^k\partial_k,\quad \delta^*=\bar{\textbf{m}}^k\partial_k\\
\kappa=-\gamma_{311},\;\sigma=-\gamma_{313},\;\lambda=\gamma_{424},\;\nu=\gamma_{422},\;\rho=-\gamma_{314},\;\mu=\gamma_{423},\;\tau=-\gamma_{312},\; \pi=\gamma_{421}\notag\\
\epsilon=\frac{1}{2}(\gamma_{341}-\gamma_{211}),\quad\gamma=\frac{1}{2}(\gamma_{342}-\gamma_{212}),\quad \alpha=\frac{1}{2}(\gamma_{344}-\gamma_{214}),\quad \beta=\frac{1}{2}(\gamma_{343}-\gamma_{213}).
\end{gather}
It should be pointed out that as a general rule - the complex conjugate of any quantity can be obtained by interchanging the indices 3 and 4 in any expression. Furthermore, the Riemann tensor can be split into a trace-free part (the Weyl tensor $C_{abcd}$, with $\eta^{ad}C_{abcd}=0$) and a trace part - given by the Ricci tensor ($R_{ac}=\eta^{bd}R_{abcd}$) and Rici scalar ($R=\eta^{ab}R_{ab}=2(R_{34}-R_{12})$). To define these the NP formalism firstly supplies five complex scalars, which completely determine the ten\footnote{This only holds in four dimensions} independent components of the Weyl tensor - $\Psi_0,...,\Psi_4$ -
\begin{gather}
\Psi_0=C_{1313}=C_{abcd}\textbf{l}^a\textbf{m}^b\textbf{l}^c\textbf{m}^d,\quad\Psi_1=C_{1213}=C_{abcd}\textbf{l}^a\textbf{n}^b\textbf{l}^c\textbf{m}^d,\quad\Psi_2=C_{1342}=C_{abcd}\textbf{l}^a\textbf{m}^b\bar{\textbf{m}}^c\textbf{n}^d\notag\\
\Psi_3=C_{1242}=C_{abcd}\textbf{l}^a\textbf{n}^b\bar{\textbf{m}}^c\textbf{n}^d,\quad\Psi_0=C_{2424}=C_{abcd}\textbf{n}^a\bar{\textbf{m}}^b\textbf{n}^c\bar{\textbf{m}}^d,
\end{gather}
and secondly, three more complex scalars and four real ones for the ten independent components of the Ricci tensor
\begin{gather}
\Phi_{00}=\bar{\Phi}_{00}=\frac{1}{2}R_{44},\quad\Phi_{01}=\bar{\Phi}_{10}=\frac{1}{2}R_{41},\quad\Phi_{02}=\bar{\Phi}_{20}=\frac{1}{2}R_{11},\quad\Phi_{11}=\bar{\Phi}_{11}=\frac{1}{4}\left(R_{43}\right),\notag\\
\Phi_{12}=\bar{\Phi}_{21}=\frac{1}{2}R_{31},\quad\Phi_{22}=\bar{\Phi}_{22}=\frac{1}{2}R_{33}.
\end{gather}
This is all that is needed in order to specify everything else - the Riemann tensor\footnote{$C_{abcd}=R_{abcd}-\left(\eta_{a[c}R_{d]b}-\eta_{b[c}R_{d]a}\right)+\frac{1}{3}R\eta_{a[c}\eta_{d]b}$ in tetrad components.}, the Ricci and Bianchi identities. These, as before, will not be presented here as they are rather long and not extremely illuminating, but an extra line will be given just to specify the components of the Maxwell field-strength tensor in terms of complex scalars, as it is needed when electromagnetic perturbations of the spacetime are investigated
\begin{align}
\phi_0=-F_{ab}\textbf{l}^a\textbf{m}^b,\quad\phi_1=-\frac{1}{2}F_{ab}\left(\textbf{l}^a\textbf{n}^b-\textbf{m}^a\bar{\textbf{m}}^b\right),\quad\phi_2=F_{ab}\textbf{n}^a\bar{\textbf{m}}^b.
\end{align}
By the above definitions it is not at all obvious why the NP-tetrad formalism should be any more special than a straightforward choice of an orthonormal tetrad. However, the real power of such a null-tetrad becomes clear once the Petrov classification of the Weyl tensor and the Goldberg-Sachs theorem have been considered. These will not be fully covered here, as detailed proofs are available in the already mentioned references, nevertheless a brief overview of the logic behind them will be presented. Clearly, the null frame (or any other tetrad frame) can be subjected to Lorentz transformations, which provide six degrees of freedom (corresponding to the six specifying parameters of the Lorentz group in 4D) to rotate the frame. These can be devised in such a way as to make a general Lorentz transformation be comprised of three types of rotations that act differently on the different tetrad basis vectors and hence on all other quantities. Moreover, in this work and in many others it is usually solutions to the vacuum Einstein equations that are investigated\footnote{With matter often introduced as a perturbation, as in the case with the massless scalar field in the previous subsection.}, in which case the Riemann curvature and Weyl tensors coincide. The latter is described by the five complex scalars introduced earlier and these are exactly the focus of the Petrov classification, which basically explores how many of them can be set to zero by a suitable orientation of the tetrad frame with the help of a Lorentz transformation. This is achieved by combining all five of them in a fourth-order equation for the parameter of one of the classes of rotations discussed just above and then looking at the possibilities in terms of the roots and Lorentz transformations. This leads to organisation of different spacetimes into five Petrov types - I, II ,III ,D and N. Remarkably, it turns out that black hole solutions of General Relativity are all of type D, which very fortunately turns out to have only one of the five Weyl scalars non-vanishing and this is $\Psi_2$. The story is not over yet, nonetheless, as by choosing the vectors $\textbf{l}$ to form a null-congruence of geodesics and referring to the Goldberg-Sachs theorem\footnote{Which applies to the Petrov type II, but leads to a corollary for type D}, it can be shown that the spin coefficients $\kappa$, $\sigma$, $\nu$ and $\lambda$ also vanish. Finally, the null geodesics in the congruence can always be chosen to be affinely parametrised which in addition also sets $\epsilon=0$. All these quite remarkable conclusions are what makes the Newman-Penrose formalism so special and the reader is encouraged to go over a detailed analysis of all this.\\
\hspace*{5mm}Moving on to what we are really interested in - perturbing the spacetime. From everything aforementioned - the most general perturbation of a type D spacetime - like Kerr-AdS or RN-AdS - will split in two parts - changes in the quantities that vanish in the unperturbed background - $\delta\Psi_0,\;\delta\Psi_1,\;\delta\Psi_3,\;\delta\Psi_4,\;\delta\kappa,\;\delta\sigma,\;\delta\lambda,\;\delta\nu$ and changes in all the rest, which do not vanish in the background (including the three complex scalars specifying the Maxwell field-strength tensor). This is worked out in excruciating detail in \cite{chandrasekhar1998mathematical} for the cases of Schwarzschild, RN and Kerr black holes. What is astonishing in all the cases is that it is possible to go on solving for the first group of quantities, listed a few sentences ago, without having to refer to any of the other perturbed variables, and successfully do so. Moreover, it turns out that the system of equations can always be reduced to a set of two equations for the Weyl scalars $\delta\Psi_0$ and $\delta\Psi_4$\footnote{$\delta\Psi_1$ and $\delta\Psi_3$ can be made to vanish by an infinitesimal coordinate rotation - this type of transformation provides four more degrees of freedom, in addition to the six due to Lorentz transformations. $\Psi_0$ and $\Psi_4$ are invariant under gauge transformations in zeroth and first order.}, which in turn allow to be separated in radial and angular parts, whereby as in the case for the massless scalar field, discussed earlier, the separation constants in both equations can be shown to be the same. Not only this, but the radial equations are complex conjugates to each other, while the angular ones are linked through a simple relation, making it sufficient to solve for only one set of them. Continuing in this fashion and fixing the normalisation of the angular solutions, the only ingredient left undetermined is the relative normalisation of the radial parts of the solution. Notwithstanding, this obstacle can also be overcome, this time with the help of the Starobinsky-Teukolsky identities, which are a small group of theorems providing a very useful set of functional transformations between the differential operators involved in the radial and angular equations. Finally, to top off all the amazing results in this computation, it has been shown by Chandrasekhar in his book\cite{chandrasekhar1998mathematical} that the rest of the perturbed quantities are also fully determined by the solutions for $\delta\Psi_0$ and $\delta\Psi_4$. This, combined with the fact that the metric components can be expressed in terms of the tetrad basis vectors \eqref{eq:MetricCompsTetrad}, implies that the most general perturbations of the metric (scalar, electromagnetic or gravitational) in the case of Schwarzschild, RN or Kerr, can be obtained from solving two separable differential equations for the two Weyl scalars $\delta\Psi_0$ and $\delta\Psi_4$ in the Newman-Penrose tetrad formalism. Fortunately, in the derivation of this result the asymptotic character of the background spacetime plays no role, as it should be, since actually solving the equations for the Weyl scalars has not been attempted yet, thus absolutely straightforwardly the above conclusions would hold for Schwarzschild-AdS, RN-AdS and Kerr-AdS. The only difference comes when one tries to solve for $\delta\Psi_0$ and $\delta\Psi_4$ and has to be cautious with what boundary conditions are imposed at spatial infinity.\\
\hspace*{5mm}In order to complete this discussion we feel that the Teukolsky master equation in its most general form for Kerr-AdS, applicable to any type of perturbation, should be given explicitly. Therefore, the Kerr-AdS metric in four dimensions, discovered by Carter\cite{carter1968hamilton} will be introduced first in the usual Boyer-Lindquist coordinates $\{\hat{t},\hat{r},\theta,\hat{\phi}\}$
\begin{align}
ds^2=-\frac{\Delta_r}{\Sigma^2}\left(d\hat{t}-\frac{a}{\Xi}\sin^2\theta\,d\hat{\phi}\right)^2+\frac{\Sigma^2}{\Delta_r}d\hat{r}^2+\frac{\Sigma^2}{\Delta_\theta}d\theta^2+\frac{\Delta_\theta}{\Sigma^2}\sin^2\theta\left(ad\hat{t}-\frac{\hat{r}^2+a^2}{\Xi}d\hat{\phi}\right)^2,\label{eq:KerrAdS}
\end{align}
where
\begin{align}
\Delta_r=\left(\hat{r}^2+a^2\right)\left(1+\frac{\hat{r}^2}{L^2}\right)-2M\hat{r},\quad\Xi=1-\frac{a^2}{L^2},\quad\Delta_\theta=1-\frac{a^2}{L^2}\cos^2\theta,\quad\Sigma^2=\hat{r}^2+a^2\cos^2\theta.
\end{align}
The solution is asymptotically AdS, as mentioned earlier, with ADM mass and angular momentum $M/\Xi^2$ and $Ma/\Xi^2$, respectively, while the event horizon is located at $\hat{r}=r_+$, where $r_+$ is the largest root of $\Delta_r$. By a suitable transformation it can be checked that the above metric is asymptotic to global ${\rm AdS}_4$ in a rotating frame with angular velocity $\Omega_\infty=-a/L^2$.\\
One way of achieving this is by first doing a slight change of variables in \eqref{eq:KerrAdS} by introducing
\begin{align}
T=\Xi\hat{t}\quad\mbox{and}\quad\chi=a\cos\theta,
\end{align}
in order to get
\begin{align}
ds^2=-&\frac{\Delta_r}{(\hat{r}^2+\chi^2)\Xi^2}\left(dT-\frac{a^2-\chi^2}{a} d\hat{\phi}\right)^2+\left({\hat{r}^2+\chi^2}\right)\left(\frac{d\hat{r}^2}{\Delta_r}+\frac{d\chi^2}{\Delta_\chi}\right)+\notag\\
+&\frac{\Delta_\chi}{(\hat{r}^2+\chi^2)\Xi^2}\left(dT-\frac{a^2+\hat{r}^2}{a}d\hat{\phi}\right)^2,\label{eq:KerrAdSv2}
\end{align}
where $\Delta_\chi=(a^2-\chi^2)(1-\frac{chi^2}{L^2})$ with the angular velocity at infinity becoming $\Omega_\infty=-a/(L^2\Xi)$, followed by another coordinates transformation:
\begin{align}
&t=\frac{T}{\Xi},\hfill \hspace*{28mm}R=\frac{\sqrt{L^2(a^2+\hat{r}^2)-(L^2+\hat{r}^2)\chi^2}}{L\sqrt{\Xi}},\notag\\
&\phi=\hat{\phi}+\frac{a}{L^2}\frac{T}{\Xi},\hspace*{15mm}\cos\Theta=\frac{L\sqrt{\Xi}\hat{r}\chi}{a\sqrt{L^2(a^2+\hat{r}^2)-(L^2+\hat{r}^2)\chi^2}}.
\end{align}
One does not need to find the metric explicitly - rather only the asymptotic behaviour, as $\hat{r}\rightarrow\infty$ (and respectively $R\rightarrow\infty$), is of interest - thus working to next-to-leading order in $\hat{r}$ (or $R$) is enough. The $t$ and $\phi$ components are straightforward to handle, while for $R$ and $\Theta$ it is easier to invert their expressions for $\hat{r}^2$ and $\chi^2$ and then proceed by brute force. The result is the global ${\rm AdS}_4$ metric \eqref{eq:globalAdS} in terms of the coordinates $\{t,R,\Theta,\phi\}$.\\
Going back to Kerr-AdS, in order to move to a non-rotating frame at infinity, one can introduce the new coordinates $\{\hat{t},\hat{r},\theta,\hat{\varphi}\}=\{\hat{t},\hat{r},\theta,\hat{\phi}+\frac{a}{L^2}\hat{t}\}$, wherein the angular velocity of the horizon with respect to an observer at spatial infinity is given by
\begin{align}
\Omega_H=\frac{a}{r_+^2+a^2}\left(1-\frac{a^2}{L^2}\right),
\end{align}
which can be easily derived by finding the equation of $\hat{\varphi}$ in terms of $\hat{t}$ on the integral curves of the horizon generating Killing vector field $\xi$\footnote{One way of doing this would be to take $\xi$ as a vector field and act on the difference between its $\hat{\varphi}$ and $\hat{t}$ components.}. The expression for the metric will not be rewritten, but it is a simple task to obtain it as it just requires the replacement of $\hat{\phi}$ in the two brackets. The rotation parameter has to be bounded by $a<L$ as is evident from the expressions for the ADM mass and energy, because for a fixed horizon radius $r_+$ they diverge in the limit $a\rightarrow L$. Furthermore, the Hawking temperature of the black hole is given by
\begin{align}
T_H=\frac{r_+}{2\pi}\left(1+\frac{r_+^2}{L^2}\right)\frac{1}{r_+^2+a^2}-\frac{1}{4\pi r_+}\left(1-\frac{r_+^2}{L^2}\right),
\end{align}
which can be used to arrive at expressions for $a$ \eqref{eq:aExtremal} and $M$ at extremality, where $T_H=0$ and $\Delta_r(r_+)=0$:
\begin{align}
a_{ext}=r_+\sqrt{\frac{3r_+^2+L^2}{L^2-r_+^2}},\quad\mbox{and}\quad M_{ext}\frac{r_+\left(1+r_+^2/L^2\right)^2}{1-r_+^2/L^2}.
\end{align}
Getting Teukolsky master equation requires the introduction of a tetrad basis - which will be provided as the extension of the original tetrad used by Teukolsky to AdS spaces - known as Kinnersly's tetrad,
\begin{align}
&\textbf{l}^\mu\partial_\mu=\frac{1}{\Delta_r}\left(\left(\hat{r}^2+a^2\right)\partial_{\hat{t}}+\Delta_r\partial_{\hat{r}}+a\left(1+\frac{\hat{r}^2}{L^2}\partial_{\hat{\varphi}}\right)\right)\notag\\
&\textbf{n}^\mu\partial_\mu=\frac{1}{2\Sigma^2}\left(\left(\hat{r}^2+a^2\right)\partial_{\hat{t}}-\Delta_r\partial_{\hat{r}}+a\left(1+\frac{\hat{r}^2}{L^2}\partial_{\hat{\varphi}}\right)\right)\notag\\
&\textbf{m}^\mu\partial_\mu=\frac{\sin\theta}{\sqrt{2\Delta_\theta\left(\hat{r}+ia\cos\theta\right)}}\left(ia\partial_{\hat{t}}+\frac{\Delta_\theta}{\sin\theta}\partial_\theta+\frac{i\Delta_\theta}{\sin^2\theta}\partial_{\hat{\varphi}}\right).
\end{align}
There is an important subtlety, worth noting, concerning the application of boundary conditions to the resulting equations. The components of the metric perturbations, which can be obtained from the solution to Teukolsky master equation by what is called the Hertz map\cite{kegeles1979constructive}, will clearly depend on the picked tetrad basis, implying that matching these with the requirements for their decay rates at $\mathcal{I}$, as discussed in \ref{sec:AAdS}, is also dependent on this choice. Unfortunately, for Kerr-AdS this has not been achieved in the aforementioned Kinnersly tetrad, but in the Chambers-Moss one\cite{chambers1994stability,dias2013boundary}, where Teukolsky equation takes on a different form than the one presented here. We will glance over this issue and hope that in the near future someone will derive the required boundary conditions. The non vanishing Weyl-scalar is given by $\Psi_2=-M(r-ia\cos\theta)^{-3}$ and we are working in vacuum. The perturbations are naturally decomposed as
\begin{align}
\Psi^{(s)}=e^{-i\omega\hat{t}}e^{im\hat{\varphi}}R^{(s)}_{lm\omega}(\hat{r})S^{(s)}_{lm\omega}(\theta),
\end{align}
where $s=0$ corresponds to scalar perturbations with $\Psi^{0}=\Psi$, $s=1$ designates electromagnetic waves with $\Psi^{(1)}=\delta\phi_0$ and $\Psi^{(-1)}=\left(-\Psi_2\right)^{-\frac{2}{3}}\delta\phi_2$, while $s=2$ denotes gravitational perturbations with $\Psi^{(2)}=\delta\Psi_0$ and $\Psi^{(-2)}=\left(-\Psi_2\right)^{-\frac{4}{3}}\delta\Psi_4$. The equation is also valid for $s=\pm\frac{1}{2}$, which is the case of massless fermions but will not be given here. With all these definitions, the radial and angular parts of the Teukolsky master equation can be presented:
\begin{gather}
\Delta_r^{-s}\partial_{\hat{r}}\left[\Delta_4^{s+1}\partial_{\hat{r}}R_{lm\omega}^{(s)}(\hat{r})\right]+\Bigg\{\frac{K_T^2-is\Delta_r'K_T}{\Delta_r}+2isK_T'-|s|(|s|-1)(2|s|-1)(2|s|-7)\frac{\hat{r}^2}{3L^2}+\notag\\
+\frac{s+|s|}{2}\Delta''-|s|\left(|s|-2\right)(4s^2-12|s|+11)\frac{a^2}{3L^2}-\hat{\lambda}^{(s)}_{lm\omega}\Bigg\}R^{(s)}_{lm\omega}(\hat{r})=0,
\end{gather}
where 
\begin{align}
K_T(\hat{r})=\omega(\hat{r}^2+a^2)-ma\left(1+\frac{\hat{r}^2}{L^2}\right)\quad\mbox{and}\quad\hat{\lambda}^{(s)}_{lm\omega}=\hat{\Lambda}^{(s)}_{lm\omega}-2ma\omega+a^2\omega^2+(s+|s|),
\end{align}
and $\hat{\Lambda}^{(s)}_{lm\omega}$ is the separation constant which will get more elaboration after the angular part of the equation has been shown:
\begin{gather}
\frac{1}{\sin\theta}\partial_\theta\left(\sin\theta\Delta_\theta\partial_\theta S^{(s)}_{lm\omega}(\theta)\right)+\left[\left(a\omega\cos\theta\right)^2\frac{\Xi}{\Delta_\theta}-2sa\omega\cos\theta\frac{
\Xi}{\Delta_\theta}+s+\hat{\Lambda}^{(s)}_{lm\omega}-\right.\notag\\
\left.-\left(m+s\cos\theta\frac{\Xi}{\Delta_\theta}\right)^2\frac{\Delta_\theta}{\sin^2\theta}-2\delta_s\frac{a^2}{L^2}\sin^2\theta\right]S^{(s)}_{lm\omega}(\theta)=0,
\end{gather}
where $\delta_s=1$ for $|s|=\{1/2,1,2\}$ and $\delta_s=0$ if $s=0$. The eigenfunctions $e^{im\hat{\varphi}}S^{(s)}_{lm\omega}(\theta)$ are the so called spin-weighed AdS spheroidal harmonics - a generalisation of their flat counterparts, with $l$ - a positive integer identified with the number of zeroes along the polar direction, which is given by the relation $l-\mbox{max}\{|m|,|s|\}$. The separation constants $\hat{\Lambda}^{(s)}_{lm\omega}$ are their associated eigenvalues and can be determined numerically, with the leading order contribution in the regime $a/L\ll1$ (which was discussed earlier in the case of the massless scalar field) proportional to $l$ and $s$ only, which will be seen later in the section on Kerr-AdS. Similar to the case of the ordinary spherical harmonics, regularity requires that $-l\leq m\leq l$ and $m\in\mathbb{Z}$. Furthermore, as mentioned before $\Phi^{(s)}_{lm\omega}(\hat{r})$ is the complex conjugate of $\Phi^{(-s)}_{lm\omega}(\hat{r})$, hence their differential equations are complex conjugates as well, while the angular solutions are related by $S^{s}_{lm\omega}(\theta)=S^{-s}_{lm\omega}(-\theta)$ and can be freely normalised by
\begin{align}
\int_{0}^{\pi}\left(S^{(s)}_{lm\omega}\right)^2d\theta=1.
\end{align}
Taking $L\rightarrow\infty$, corresponding to a vanishing Cosmological constant, reduces the above equations to the case of the flat Kerr solution. Moreover, in the limit of $a\rightarrow0$ the radial and angular equations take on the form appropriate for Schwarzshild-AdS (or Schwarzschild if $L\rightarrow\infty$ has already been taken), from where the RN-AdS form of the equations can be deduced by changing $\left.\Delta_r\right|_{a\rightarrow0}$ to $\left.\Delta_r\right|_{a\rightarrow0}+e^2$, where $e^2=\sqrt{Q^2+P^2}$, with $Q$ and $P$ representing the electric and magnetic charges of the black hole, respectively. In addition, setting $M=0$ in the Schwarzschild equations produces the global ${\rm AdS}_4$ ones. Having the radial and angular parts of Teukolsky master equation readily available, in order to study some type of perturbations of Kerr-AdS (or Schwarzschild- or RN-AdS), one just needs to take the appropriate value for $s$ and start solving. The approach is the same as for the massless scalar field - a complete analytical solution is not known currently, but a perturbative expansion near the horizon and at large radial distances, where attention has to be paid in applying the correct boundary conditions to ensure that $\mathcal{I}$ acts as a reflecting wall, combined with an asymptotic matching procedure in an intermediate region, with the same assumptions as before on the parameters of the black hole, provides a very good approximation in the regime of small black holes. Afterwards, a numerical analysis can be performed in the whole phase space in order to confirm the perturbative results and investigate what happens for large black holes. The conclusions of such types of investigations of the QNMs and superradiant modes of RN-AdS and Kerr-AdS will be presented in the next two sections, where a particular focus will be paid to the superradiance and its effect on the stability of the spacetimes under question, but as one might expect the inequalities \eqref{eq:ineqRN} and \eqref{eq:ineqKerr} for the real and imaginary parts of the frequencies of the perturbative modes will show up again.
\section{Superradiance in Reissner-Nordstr\"om-AdS and its stability}\label{sec:sec4}
Reissner-Nordstr\"om metrics are not expected to represent black holes of particular importance to astrophysics, due to the charge neutrality of the universe, which implies that large charge imbalances are unlikely to occur. Furthermore, a charged black hole would definitely attract particles of opposite charge and will eventually lose most of its charge. Nevertheless the RN-AdS metric is a manageable toy model for superradiance (which does not occur in Schwarzschild-AdS) and from the fairly simple arguments put out in subsection \ref{sec:SimpleEx}, one might expect that it should be possible to make some analogies between RN-AdS and Kerr-AdS. There definitely are similarities between the phenomenon in the two spacetimes, although this might be more due to the fact that it is the same problem being investigated. As it turns out, translating the results for charged black holes into conclusions for rotating black holes is clearly not straightforward.\\
\hspace*{5mm}The RN-AdS metric in 4D is given by
\begin{align}
ds^2=-\frac{\Delta}{r^2}dt^2+\frac{r^2}{\Delta}dr^2+r^2(d\theta^2+\sin^2\theta\,d\phi^2),
\end{align}
where (taking the magnetic charge $P=0$)
\begin{align}
\Delta=\frac{r^4}{L^2}+r^2-2Mr+Q^2,
\end{align}
while the Hawking temperature and the mass (in terms of the horizon radius) of the black hole are evaluated to be
\begin{align}
T_H=\frac{1}{4\pi r_+}\left(1-\frac{Q^2}{r_+^2}+\frac{3r_+^2}{L^2}\right)\quad\mbox{and}\quad M=\frac{1}{2}\left(r_++\frac{r_+^3}{L^2}+\frac{Q^2}{r_+}\right),
\end{align}
which as before can be used to get the parameters at extremality, leading to
\begin{align}
Q_{ext}=r_+\sqrt{1+\frac{3r_+^2}{L^2}}.
\end{align}
As the black hole is charged and posses no angular momentum, the simplest way of achieving superradiance is through a massless charged scalar field which was investigated analytically in \cite{uchikata2011quasinormal} and then fully numerically in \cite{bosch2016nonlinear}. It should be noted that all the considerations laid out in the subsection on scalar fields \ref{subsec:Scalar} apply here - one just needs to give the field a charge, which we will take to be designated by $e$ from here onwards. Moreover, a complete phase space diagram of static, charged, asymptotically AdS solutions in 5D in terms of the charge $Q$ and the mass $M$ ($x$- and $y$-axis, respectively) of the solution is available and was obtained in the microcanonical ensemble in \cite{dias2012hairy}. It consists of static charged solitons, RN-AdS black holes and `hairy' black holes. Technically, there are more possible solutions - excited solitons or excited hairy black holes - but these should not be important as long as the charge of the scalar field is not very large. The first of the three solutions is basically a static blob of charged condensate with zero entropy in global ${\rm AdS}_5$, which for a given value of $e$ is entirely determined by its charge, whereby in the limit of the latter being infinitesimally small, the condensate reduces to the lowest energy linear perturbation of ${\rm AdS}_5$ by the scalar field. Skipping the second solution, as it is well-known, the third one represents a black hole, which is not entirely depleted of charge with a charged scalar condensate around it\footnote{It can be said that the condensate is an orbiting hair, but it should not be forgotten that the orbit is static - that is - nothing rotates around the black hole, as there is no angular momentum in the system.}. Depending on the value of $e$\footnote{The exact numerical values can be looked up in \cite{dias2012hairy}.} there are three different possible configurations of the phase space. For small charges of the scalar field there are only RN-AdS black holes and charged solitons, with the former always being the dominant phase from an entropy point of view, while the latter exist only up to some finite value of the charge $Q_{crit}$ and there are no instabilities in the system. However, in an intermediate range of values for the charge $e$, the RN-AdS black holes become unstable near extremality, with a condition on their charge $Q$ as a function of the scalar field one $e$. As expected from perturbative analysis, at the onset curve of this instability, the hairy black hole solutions branch off. The numerical construction in \cite{dias2012hairy} has shown that they do exist below the extremality curve for RN-AdS black holes and are the thermodynamically preferred solution whenever they appear. Solitons are also present up to some finite charge $Q_{crit}$, but for a given mass and charge are never the dominant stable solution in the phase diagram. Finally, in the third possible regime, for $e$ higher than some numerically found critical value (but not analytically justified yet), the RN-AdS black holes are unconditionally unstable near extremality. Moreover, this time the solitonic solution always exist, with masses below the extremal curve and it represents the ground state of the system for black hole charges below a certain transitioning value $Q_{c_2}$, whereas for $Q>Q_{c_2}$ it is again the hairy black holes that become thermodynamically favoured, but this time reducing in their zero mass limit to an infinite temperature soliton (they still branch off at the onset of instability). All these considerations have been deduced for the five-dimensional static charged vacuum solution of Einstein equations with negative cosmological constant, but one expects that the behaviour in four dimensions will not be qualitatively different. It should also be mentioned that the analysis in \cite{dias2012hairy} takes into account two types of mechanisms leading to instability - the first applies for black holes with radius much larger than the AdS curvature and is the result of the violation of the Breitenl\"ohner-Freedman (BF) bound of the near horizon extremal geometry (with topology ${\rm AdS}_2$) by the mass of the scalar field.\\
A detailed discussion will not be presented, but an intuitive description of the BF bound can be obtained in a simple way. Take a massive scalar field in pure ${\rm AdS}_d$ space and examine its Klein-Gordon equation $\left(\nabla_\mu\nabla^\mu-\mu^2\right)\Phi=0$ - by decomposing the scalar field according to the symmetries of the spacetime, as usual, and carrying out certain algebraic manipulations, a Schr\"odinger-like equation is derived. By consequently studying its potential, a condition on the stability of the solutions can be obtained, which gives the BF bound on the mass of the scalar field. One can perform a similar procedure for a massive scalar field in an asymptotically ${\rm AdS}_d$ spacetime that satisfies the Klein-Gordon equation $\left(\nabla_\mu\nabla^\mu-\mu^2\right)\Phi=0$, whereby doing a Frobenius analysis at infinity and requiring that the powers of the resulting coefficients of the solution are real provides a bound on the mass of the scalar field. Afterwards, looking at an extremal, asymptotically AdS black hole solution and considering the near horizon region, where the topology contains ${\rm AdS}_2$, whose BF bound is above the one of the background AdS space, it is realised that for scalar fields with masses between the two BF values, the near horizon geometry will be unstable, while the asymptotic space will not be.\\ The second kind of instability is due to superradiance and is applicable in the case of small black holes and is what interests us mainly. As already mentioned, in four dimensions the first detailed investigation of the superradiant modes and the instability of RN-AdS is carried out in \cite{uchikata2011quasinormal}, where the authors first perform the analytical analysis as outlined in \ref{subsec:Scalar} and find that the frequencies\footnote{The authors work with a few different frequency definitions, differing by a constant that is coming from the potential difference due to charge of the black hole - we have listed all the results in terms of the original $\omega$, used in the decomposition of the scalar field. The superradiance condition itself does not depend on that constant as one might expect.} are quantised as follows
\begin{gather}
\mbox{Re}(\omega)=\frac{2n+l+3}{L}\\
\mbox{Im}(\omega)=-\sigma_0\frac{(l!)^2(l+2+n)!2^{l+3}(2l+1+2n)!!}{(2l+1)!(2l)!n!(2l-1)!!(2l+1)!!(2n+3)!!}\frac{(r_+-r_-)^{2l+1}}{\pi L^{2l+2}}\prod_{k=1}^{l}(k^2+\sigma_0^2),
\end{gather}
where
\begin{align}
\sigma_0=\left(\omega_0-e\frac{Q}{r_+}\right)\frac{r_+^2}{r_+-r^-}\quad\mbox{with}\quad\omega_0=\frac{2n+l+3}{L}=\mbox{Re}(\omega),
\end{align}
where $\omega_0$ represents the QNM frequencies of the pure ${\rm AdS}_4$ spacetime, which are derived from Teukolsky equation in \cite{dias2013boundary}, with $n$ a non-negative integer called the radial overtone, which gives the number of nodes along the radial direction (of the radial eigenfunction). Clearly, $\sigma_0$ determines the sign of the imaginary part of the frequency, which in turn dictates whether the mode is superradiant or quasinormal. For $\sigma_0<0$ one gets Im$(\omega)>0$ and hence the mode is exponentially growing in time, whereas for $\sigma_0>0$, when Im$(\omega)<0$, the modes are quasinormal. As seen from the formula for $\sigma_0$, these statements are equivalent to
\begin{align}
\mbox{Re}(\omega)-e\frac{Q}{r_+}<0\;\Rightarrow\;\mbox{Im}(\omega)>0,\quad\mbox{and}\quad\mbox{Re}(\omega)-e\frac{Q}{r_+}>0\;\Rightarrow\;\mbox{Im}(\omega)<0
\end{align}
which is exactly the condition for superradiance that was presented earlier in section \ref{subsec:Scalar}. The authors of \cite{uchikata2011quasinormal} also provide a numerical investigation that supports their claims based on the perturbative analysis and show that indeed the imaginary part of the frequency changes sign, when the superradiant condition is met, however they do not do it at the full non-linear level, which would enable one to accurately say what is the endpoint of the superradiant instability. Fortunately, this is done in \cite{bosch2016nonlinear}. The simulations that have been carried out by the authors confirm the analytical results described just above as long as the perturbation remains small - which is expected, as for significant perturbations the non-linear effects and the backreaction on the spacetime become important. Furthermore, it is found that for small charges $e$ of the scalar field there are no unstable modes present - agreeing with what is discovered in \cite{dias2012hairy}. In the presence of superradiant modes, both mass and charge are extracted from the black hole by them, with the ratio between the two depending on the initial value of $e$ - the larger it is, the more charge and less mass is drained. The resulting charged scalar hair `orbits' the black hole with its distance from the black hole increasing with increasing $e$. The evolution of this instability proceeds in the following way - firstly, based on the initial data there will be a mix of QNMs and superradiant modes. The former will quickly decay, whereas the latter will steadily grow with time. Nonetheless, while the extraction is in progress, the charge $Q$ of the black hole will decrease, while the horizon radius $r_+$ will be increasing, consistent with the second law of black hole mechanics, meaning that the superradiance condition will be getting more and more stringent. This implies that gradually the superradiant modes (starting from large $n$) will cease being such and turn into QNMs and eventually decay and get absorbed by the black hole, restoring a bit of its mass and charge, but not enough to compensate for the extraction (the fundamental $n=0$ mode is the most effective at extracting). This goes on until only the fundamental mode $n=0$ is left - neither growing, nor decaying (this is seen from the simulations) - as a charged scalar condensate `orbiting' the black hole. Therefore, the endpoint of the instability, due to superradiance, for RN-AdS is at a hairy black hole, whereby the hair consists in a charged scalar condensate `orbiting' the black hole at a distance - the four-dimensional equivalent of the hairy black holes constructed in \cite{dias2012hairy}. A few remarks are in order here. After the fundamental mode has reached zero growth rate, it starts oscillating harmonically, which implies that the scalar field stress-energy tensor becomes time-independent and there are no more changes in the metric. Moreover, it is observed that the higher the value of $e$ is, the faster the whole evolution proceeds and as already mentioned much more charge than mass is extracted and the scalar hair condensates further away from the black hole. In the limit of very large scalar charge $e$ it is expected that the resulting configuration will be a Schwarzschild-AdS black hole with a scalar condensate very far away. Finally, we mentioned before that one might be tempted to make analogies between what happens in RN-AdS and what might happen in Kerr-AdS, by naively looking at the conditions for superradiance derived earlier (the one for Kerr-AdS is also confirmed in the literature)
\begin{align}
\mbox{Re}(\omega)-e\frac{Q}{r_+}<0\quad\mbox{and}\quad\mbox{Re}(\omega)-m\Omega_H<0.
\end{align}
This, unfortunately is not possible, due to the fact that in the first case the scalar field charge $e$ is held fixed at a given value, whereas for Kerr-AdS $m$ is allowed to take on any integer value. corresponding to the active superradiant modes, resulting in more complicated dynamics for the instability. Nevertheless, one might speculate that similar to the situation just discussed, the condition for superradiance will be getting stronger and stronger until only a single superradiant mode is left excited. An evolution at the fully non-linear level has not been carried out for Kerr-AdS yet, but there are a lot of results that point to very interesting possibilities for the endpoint of its instability, as it will be shown in the next section.
\vspace{-1mm}
\section{Kerr-AdS, Superradiance and the problem with instability}\label{sec:sec5}
\vspace{-1mm}
The previous section started with the remark that Reissner-Nordstr\"om black holes are not particularly relevant to astrophysical observations, but a similar comment can be made about Kerr-AdS solutions, as according to cosmological observations our Universe is almost flat. Nonetheless, as eluded to earlier, one can imagine a massive scalar field creating a trapping potential at a distance comparable to the radius of curvature of AdS, which would make it possible to compare the two situations. This is not the only reason why Kerr-AdS is attracting attention recently (as is evident from the growing number of papers on the topic) - from what has been done up to now in terms of research in the area it is currently not clear what the endpoint of its superradiant instability is. This is a rather delicate question with regards to the cosmic censorship conjectures as will become clearer soon, as we present the work that has been carried out on the subject until now.\\
\hspace*{5mm}The metric for Kerr-AdS was given in \ref{subsec:NPFormalism} and it will not be presented here again. The first analytical study of its QNM and superradiant modes was carried ot in \cite{cardoso2004small}. The type of perturbation considered was again a massless scalar field (uncharged) and the path of the analysis was similar to the one laid out in section \ref{subsec:Scalar} and subsequently repeated in the previous section on RN-AdS. The calculation starts from the wave equation for the field $\nabla_\mu\nabla^\mu\Phi=0$ and proceeds through the same decomposition of the field according to the background symmetries and then finishes with an asymptotic matching procedure in a zone where the near-horizon and far away regions overlap for the range of parameters $a/r_+\ll1$ and $r_+\omega\ll1$, corresponding to small black holes. Unfortunately, the authors did not impose the appropriate reflecting boundary conditions at $\mathcal{I}$ that will preserve the boundary metric and hence the asymptotic AdS structure. This was corrected in \cite{uchikata2009scalar}, where a numerical study of the problem was also supplemented in order to confirm the perturbative analysis and show that indeed small Kerr-AdS black holes are unstable to superradiant modes. We pause to say that a detailed derivation of the required boundary conditions at $\mathcal{I}$ for a general perturbation, corresponding to the definition of asymptotically AdS, given in section \ref{sec:AAdS}, is available in \cite{dias2013boundary}. Back to the scalar field perturbations - as for RN-AdS the analytically determined expression for the frequencies of the modes will be displayed, as found in \cite{uchikata2009scalar}:
\begin{align}
\omega&=\omega_0+i\delta,\quad\mbox{where}\quad\omega_0=\frac{2n+l+3}{L}=\mbox{Re}(\omega)\\
\delta&\approx-\sigma\left(\omega_0-m\Omega_H\right)\frac{(r_+^2+a^2)(r_+-r_-)^{2l}}{\pi L^{2(l+1)}},
\end{align}
where
\begin{gather}
\sigma=\frac{(l!)^2(l+2+n)!}{(2l+1)!(2l)!n!}\frac{2^{l+3}(2l+1+2n)!!}{(2l-1)!!(2l+1)!!(2n+3)!!}\prod_{k=1}^{l}\left(k^2+4\varpi^2\right),\\
\mbox{with}\quad\varpi=\left(\omega_0-m\Omega_H\right)\frac{r_+^2+a^2}{r_+-r_-},
\end{gather}
The results look of qualitatively the same form as for RN-AdS - the real part of the frequency is again equal to the normal modes of global ${\rm AdS}_4$, with $n$ a non-negative integer that denotes the radial overtone as before and the structure of the imaginary part is very similar. Its sign is determined by what is called the superradiant factor $\varpi$ (through the combination $\omega_0-m\Omega_H$) and the superradiance condition takes the form
\begin{align}
\mbox{Re}(\omega)-m\Omega_H<0\;\Rightarrow\;\mbox{Im}(\omega)>0,\quad\mbox{and}\quad\mbox{Re}(\omega)-m\Omega_H>0\;\Rightarrow\;\mbox{Im}(\omega)<0,
\end{align}
where the former implies that the mode is exponentially growing in time, hence superradiant, while the latter designates a QNM. As we already mentioned, the numerical results presented in \cite{uchikata2009scalar} confirm the above formulae and show that indeed there are superradiant modes present in the spectrum of massless scalar field perturbation of Kerr-AdS, depending on the value of the rotation parameter. Unfortunately, the authors do not manage to clarify the nature of this dependence - but the numerics suggest that for smaller black holes faster rotation implies more superradiance (possibly up to some critical value of $a$). As briefly remarked in the introduction of the section, there are no fully-nonlinear simulations performed for Kerr-AdS yet, thus we move on to another type of spacetime probing, which will be followed by a tour of the possible evolution of the system.\\
\hspace*{5mm}The analogue of the aforementioned analysis in the case of gravitational perturbations is performed in \cite{cardoso2014holographic} (The task was firts attempted in \cite{cardoso2006classical}, but unfortunately with incorrect boundary conditions at infinity). The approach to the perturbative calculations is very similar - it just starts from the Teukolsky equation for the case of gravitational perturbations - $s=\pm2$. Afterwards, approximating solutions in terms of hypergeometric functions are obtained in the near-horizon and far regions and then matched in an overlapping zone as before, while being cautious to impose the correct boundary conditions as prescribed by \cite{dias2013boundary}. The quantised spectrum of the perturbations' frequency will not be shown here, as it is much longer than the ones given before and contains hypergeometric functions, which makes it rather less illuminating. It is important to note that due to the restriction on $l$ that was given in the remarks following Teukolsky master equations, the smallest value it can take is $l=|s|=2$, implying that the Teukolsky formalism misses out two of the modes of the perturbations. Fortunately, in \cite{dias2013algebraically} it was proven that these modes only shift the mass and angular momentum of the solution, hence only correspond to deformations within the Kerr-AdS family. The perturbation sector under question is separated in two - scalar and vector gravitational perturbations and it is to be noted that the authors of \cite{cardoso2014holographic} concentrate on modes of the type $l=m$, but this should not have any effects on the qualitative results presetned. The analytical investigation indicates that the real part of the frequency in both sectors is very close to the values of the corresponding normal modes in global ${\rm AdS}_4$, which are given in \cite{dias2013boundary} (produced by solving Teukolsky equation with $a=0$ and $M=0$ for $s=2$). The imaginary parts can be either calculated numerically or expanded in series of the rotation parameter and horizon radius, both divided by the AdS curvature radius, in accordance with the parameter regime for Kerr-AdS that we introduced earlier and that is used in the paper for the analytical computations - $r_+/L\ll1$ and $a/L\ll1$. The expressions for Im$(\omega)$ resulting from these series expansions show that for $a=0$ the modes are always quasinormal and thus decaying, agreeing  with results from Schwarzschild-AdS, where there are no superradiant modes\cite{dias2013boundary}. Furthermore, it is seen that for Im$(\omega)=0$ one gets Re$(\omega)-m\Omega_H\approx0$, while for Re$(\omega)-m\Omega_H>0$ the modes are damped with Im$(\omega)<0$ and if Re$(\omega)-m\Omega_H<0$, then Im$(\omega)>0$, indicating superradiant modes and confirming again the familiar condition for superradiance. Interestingly, the authors find that the Im$(\omega)$ increases with faster rotation, similar to the results of \cite{uchikata2009scalar}. The consequently presented numerical investigation of the modes, apart from confirming the analytical results, demonstrates that they posses a few interesting properties. Firstly, plotting the onset of the superradiant instability (where the imaginary part vanishes and $\omega=m\Omega_H$) as a function of the angular velocity and the (gauge invariant) radius of the black hole (that is - a contour plot as a function of $\Omega_H/L$ and $R_+/L$, where $R_+=\frac{\sqrt{r_+^2+a^2}}{\sqrt{\Xi}}$), it is discovered that all the onset curves lie above the line $\Omega_HL=1$, which was first conjectured in \cite{kunduri2006gravitational}, but in the limit of $R_+/L\rightarrow\infty$ - approach it gradually (in a different way for scalars and vectors). Furthermore, in the scalar sector, for small horizon radii, the larger $l=m$ is, the lower the onset curve of the mode starts as a function of the rotation - that is for small black holes the $l=m=2$ mode is the last to go unstable, as all modes with higher $l=m$ numbers will turn on at a lower rotation parameter. However, at larger radii, things seems to start reversing and the onset curves of the modes begin to cross, such that there are regions where the $l=m=2$ mode will switch on at a lower angular velocity than the $l=m=3$ mode, for example. Nonetheless, it should be noted that in the limit of $l=m\rightarrow\infty$, the corresponding mode will be an almost horizontal line, infinitesimally close to $\Omega_HL=1$, hence these modes will always be the first to become unstable. Lastly, all the modes asymptote to the $\Omega_HL=1$ line as the black hole radius goes to infinity. For the gravitational vector modes the picture is rather different - the first modes to go unstable are still the largest ones in terms of the numbers $l=m$, but this time there are no crossings whatsoever between the onset curves. Moreover, all the onset curves end at the extremality curve for Kerr-AdS, whereby modes with higher $l=m$ end up reaching it for even larger radii, such that they are slowly approaching the $\Omega_HL=1$ line (as the extremality curve is asymptotic to it). It can be deduced that the $l=m\rightarrow\infty$ mode will only reach the extremal curve in the limit $R_+/L\rightarrow\infty$, where it should also asymptote to the $\Omega_HL=1$ line (becoming again almost horizontal). Finally, two remarks regarding both sectors - it is observed in the numerical data that for a black hole of a fixed size the highest growth rate for a superradiant mode is always close to extremality, with this being much more the case for vector modes (probably due to the fact that their onset curves end up at the extremality curve). Secondly, the strength of the gravitational perturbations seems to be higher than that of the massless scalar field up to a few orders of magnitude in some cases.\\
\hspace*{5mm}Even though it was argued that the RN-AdS model does not allow for a straightforward generalisation to rotating black holes, looking at the story there, one might expect that, in the current scenario, at the onset of superradiance there might be a new family of stationary black holes merging or bifurcating with Kerr-AdS. This was actually proposed for the first time in \cite{kunduri2006gravitational}, based on the observation that the zero mode corresponding to the onset curves -  $\omega=m\Omega_H$ and Im$(\omega)=0$ - is invariant under the horizon-generating Killing vector field $k=\partial_{\hat{t}}+\Omega_H\partial_{\hat{\varphi}}$. Hence it might be reasonable to expect the existence of black holes with a single helical KVF $k=\partial_{\hat{t}}+\Omega_H\partial_{\hat{\varphi}}$, which are neither time-symmetric, nor axisymmetric. Such a type of black holes, coupled to a matter field, were constructed perturbatively and fully numerically in the case of five-dimensional AdS background and a scalar field perturbation in \cite{dias2011black}, while the formulation as a solution to the Einstein equation with negative cosmological constant was accomplished in \cite{dias2015black}, numerically. This achievement is similar in nature to what was presented above in the situation of RN-AdS from \cite{dias2012hairy}, with the notable difference that these single KVF black holes represent a second unique solution in the system under question, together with the Meyers-Perry-AdS black holes, which are the five-dimensional generalisation of the Kerr-AdS solution\footnote{The uniqueness theorems are not violated, as the helical KVF is generating the horizon, thus is normal to it, which is in contrast with the assumptions of the theorems.}. Likewise, in this configuration the solitons are replaced by what are called rotating boson stars - smooth horizonless geometries with harmonic time dependence, parametrised by the amplitude of the scalar field (instead of its charge, as for RN-AdS), whereas the hairy black holes\footnote{Which are the single KVF black holes} have decided to change hairstyles and have opted for a chargless rotating scalar condensate. We are not going to investigate this phase space in great detail, instead the focus will be shifted towards four dimensions and gravitational perturbations (they were also shown to be stronger). Similar constructions have been devised both perturbatively and numerically in \cite{dias2015black,dias2012gravitational,horowitz2014geons}. The analogue of the charged solitons from the previous section and the aforementioned rotating boson stars are the so called geons - blobs of gravitational energy with harmonic time dependence - smooth and horizonless - they are the single mode, non-linear generalisations of some of the linearised gravitational perturbations of global ${\rm AdS}_4$ and posses helical symmetry. They have been analysed both analytically and numerically in \cite{dias2012gravitational,horowitz2014geons} with relevance to the stability of global AdS\footnote{Interestingly, global AdS is linearly stable, but non-linearly unstable due to a high number of resonances between modes, which are equidistantly spaced in the linearised theory. This will not be discussed here, but the above cited papers are a good read on the topic.}. On the other hand, the four-dimensional single Killing field black holes in this purely gravitational set up are first investigated analytically in \cite{cardoso2014holographic}, where their thermodynamic properties are derived to leading order, and then numerically in \cite{dias2015black} where they got named black resonators by the authors. This is due to the fact that these single KVF black holes branch off at the onset of superradiance for a specific superradiant mode in Kerr-AdS and thus select out its particular frequency, meaning that they are not unstable to perturbations by that mode. It is also proposed by the authors that the definition of stationary should be extended to include these solutions as well, even though they do not posses a timelike KVF, as they are still periodic in a sense, due to their helical KVF. Nevertheless, they are still unstable to modes with higher $m$ numbers as argued in the paper and as mathematically proven by the results in \cite{green2015superradiant} (with the slight caveat on this result, as elucidated earlier in \ref{sec:SimpleEx}), since the single KVF (which also generates the horizon) is not everywhere timelike, thus implying the existence of an ergoregion. This can also be expected from the results of \cite{cardoso2014holographic}, presented above, for small Kerr-AdS black holes, because it turns out that small black resonators can be approximated by small Kerr-AdS BHs centered at a geon. This is an interesting construction and the idea behind the perturbative formulation of the black resonators. The Kerr-AdS black hole is placed at the core of the geon and the angular velocities of the two are matched\footnote{So that there are no unwanted fluxes across the horizon.}, whereby the former controls the entropy and the temperature of the resulting object (geons have zero entropy and undefined temperature), whereas the latter is responsible for the single helical KVF nature of the resonators. In the limit of zero size they become geons with the picked out frequency corresponding to a normal mode of global ${\rm AdS}_4$. Henceforth, it can be said that the black resonators connect the onset of superradiance in Kerr-AdS to the horizonless geons. There are a few remarks, though, that have to be made. Firstly, the above analysis is done only for a single mode $l=m=2$, but other $l=m$ modes are expected to behave qualitatively the same, while $l\neq m$ ones are subject to investigations at the moment. Secondly, the black resonators have higher entropy than the Kerr-AdS black holes, thus whenever the two solutions coexist, the former will be favoured entropically (at the same asymptotic charges $E$ and $J$, that is). Furthermore, for a fixed energy $E$ and angular momentum $J$ the entropy of the black resonators is an increasing function of $m$, therefore progression towards superradiant modes with higher azimuthal numbers is preferred. These last two results should play an important role in the evolution of the superradiant instability. Starting with some initial data for Kerr-AdS with a mixture of QNMs and superradiant modes, the former will quickly decay, whilst the latter will go on bouncing back and forth between the horizon and the conformal boundary, extracting mass and angular momentum from the black hole on every cycle. This clearly leads to the decrease of $\Omega_H$ and should eventually result in a black hole with lumpy gravitational hair that is co-rotating with the black hole, which is invariant under only a single helical KVF - $k=\partial_{\hat{t}}+\Omega_H\partial_{\hat{\varphi}}$ - a black resonator that has branched off at the onset of superradiance. However, this cannot be the endpoint of the instability, as most likely during the evolution higher $m$ superradiant modes will be activated and while the newly formed black resonator is stable against the mode, from whose onset curve it had emerged, it is still unstable to perturbations with higher azimuthal numbers $m$. One might assume then that the system will continue evolving towards configurations with higher and higher $m$ - perhaps a mixture of a black resonator (or maybe a few) and co-rotating gravitational hair. In the limit of $m\rightarrow\infty$, as was already explained, the onset curve becomes a horizontal line, infinitesimally close to $\Omega_HL=1$ with all other onset curves lying above it, hence a configuration that is stable against this limiting mode will be stable against all other modes. Therefore the superradiance phenomenon will cease and one might expect that the resulting black hole will be a limiting black resonator with $m\rightarrow\infty$. There is a slight caveat however - it was explained earlier that in their zero size limit black resonators should represent geons - but in \cite{niehoff2015towards} the authors argue on the basis of perturbation theory and supersymmetry that such a geon does not exist, as it should be a minimum energy solution to the vacuum Einstein equations with negative cosmological constant. Unfortunately, in a supersymmetric setting AdS is the only such solution. This argument relies on the assumption of smoothness but it is also possible to envisage, as the endpoint of the instability, a limiting black resonator, whose zero size limit is a singular geon, which sticks well with the idea that singularities should be enclosed by event horizons. It turns out, though, that in perturbation theory the geons' curvature gets smaller and smaller with increasing azimuthal number $m$, thus ruling out the singularity scenario for the limiting geon. Of course, it should not be forgotten that these arguments are not backed by a full non-linear evolution of the system as in the case of RN-AdS, thus there might be other factors that will come into play as the system evolves. Finally, there is one more point worth mentioning. The numerical investigation in \cite{cardoso2014holographic} confirmed that only black holes with $\Omega_HL<1$ are stable to superradiance, implying that a logical expectation will be that single KVF black hole solutions with $\Omega_HL<1$ will be the endpoint of the superradiant instability. Unfortunately, in the very similar in nature configuration in five dimensions, which was analysed in \cite{dias2011black}, none of the fully numerically constructed single KVF black holes has $\Omega_HL<1$.\\
\hspace*{5mm}Summarising the above discussion, without a full numerical simulation, it seems that there is no regular solution that comes out at the endpoint of the superradiant instability. The system either settles down to a singularity in a finite time - violating the weak cosmic censorship, or it goes on evolving indefinitely towards configurations with even higher azimuthal number $m$ and therefore also higher entropy. This implies that eventually it would be necessary to consider physics on such small scales that the effects of quantum theory might become important, which is not what is expected from the point of view of the strong cosmic censorship, since the initial system was well defined classically.
\section{Conclusion}
In this essay we examined the phenomenon of superradiance in asymptotically AdS spacetimes, giving priority to its effect on the stability of the involved spaces. Due to its timelike boundary at spatial infinity, AdS provides us with a natural way of working in a confining box with reflecting walls, given that the correct boundary conditions  at spatial infinity - keeping the boundary metric fixed - are defined. In this set up one only needs to take advantage of the well-established Newman-Penrose-Teukolsky formalism in order to study perturbations of any type in the given spacetime by directly going on solving the Teukolsky master equation. This has been done for a plethora of configurations and we presented the results for charged (RN-AdS) and rotating (Kerr-AdS) black holes, with the obvious absence of Schwarzschild-AdS, because it does not posses any superradiant modes. Even though the former two spacetimes share many similarities in their QNM and superradiant spectra, the little difference between the condition for superradiance in both cases -  $\mbox{Re}(\omega)-q\frac{Q}{r_+}<0$ and $\mbox{Re}(\omega)-m\Omega_H<0$ - that is, the fixed value of the charge of the external perturbation $q$ for RN-AdS, versus the freedom of the azimuthal number $m$ to take on any integer value in Kerr-AdS - leads to conceptually different outcomes. While for RN-AdS a fully non-linear evolution of the system has confirmed that there is an endpoint for the superradiant instability at a black hole with static charged scalar condensate around it, for Kerr-AdS this seems unlikely due to conjectured progress of the system towards configurations with even higher $m$ modes. According to the presented in this work research in the area, at the onset of superradiance in Kerr-AdS a second stationary\footnote{With a discussion in the main text on its meaning in this case.} solution branches off, which represents the so called black resonators - black holes with a single helical KVF that is also a generator of the horizon and which are connected in their zero-size limit to smooth horizonless solutions of the vacuum Einstein equations in AdS. Unfortunately, by a mathematical result in \cite{green2015superradiant}, the spacelike nature of the single KVF in some regions implies the existence of an ergoregion, thus rendering the resonators unstable to superradiance as well. By naively following one's nose towards the limit of $m\rightarrow\infty$ one reaches the conclusion that what might be the endpoint of the instability in the form of the limiting black resonator seems to be not well-defined, as its limiting geon is proven not to exist. The conclusion that is drawn from the situation is that there are two possibilities - either the instability leads to a singularity, which violates the weak cosmic censorship, or the system evolves towards configurations which require considerations at even smaller scales, making it necessary to take quantum effects into account - going against the spirit of the strong cosmic censorship, as the system that was started from is classically well-defined. All this conclusions were derived on the basis of perturbation thery, as a fully non-linear simulation in the case of Kerr-AdS has not been carried out yet, but the above propositions only make it more exciting until the complete answer is uncovered.
\bibliographystyle{unsrt}
\bibliography{Essaybib}
\end{document}